\documentclass[aps,twocolumn,showpacs,preprintnumbers,amssymb]{revtex4}

\newcommand{\be}{\begin{eqnarray}}
\newcommand{\ee}{\end{eqnarray}}

\newcommand{\eins}{\mbox{$1 \hspace{-1.0mm}  {\bf l}$}}
\def\bea{\begin{eqnarray}}
\def\eea{\end{eqnarray}}
\def\C{\hbox{$\mit I$\kern-.7em$\mit C$}}
\def\N{\hbox{$\mit I$\kern-.3em$\mit N$}}

\def\tr{{\rm tr}}

\usepackage{dcolumn} 
\usepackage{bm} 

\usepackage{epsfig}

\begin{document}

\title{Entanglement properties of multipartite entangled states under the influence of decoherence}

\author{M. Hein$^{1}$, W. D\"{u}r$^{1}$ and H.-J. Briegel$^{1,2}$}

\affiliation{
$^1$Institut f\"ur Theoretische Physik, Universit\"at Innsbruck, A-6020 Innsbruck, Austria.\\
$^2$Institut f\"ur Quantenoptik und Quanteninformation, \"Osterreichische Akademie der Wissenschaften, A-6020 Innsbruck, Austria.}

\date{\today}

\begin{abstract}
We investigate entanglement properties of multipartite states under the influence of decoherence. We show that the lifetime of (distillable) entanglement for GHZ--type superposition states decreases with the size of the system, while for a class of other states --namely all graph states with constant degree-- the lifetime is independent of the system size. We show that these results are largely independent of the specific decoherence model and are in particular valid for all models which deal with individual couplings of particles to independent environments, described by some quantum optical master equation of Lindblad form. For GHZ states, we derive analytic expressions for the lifetime of distillable entanglement and determine when the state becomes fully separable. For all graph states, we derive lower and upper bounds on the lifetime of entanglement. The lower bound is based on a specific distillation protocol, while upper bounds are obtained by showing that states resulting from decoherence in general become non--distillable or even separable after a finite time. This is done using different methods, namely (i) the map describing the decoherence process (e.g. the action of a thermal bath on the system) becomes entanglement breaking, (ii) the resulting state becomes separable and (iii) the partial transposition with respect to certain partitions becomes positive. To this aim, we establish a method to calculate the spectrum of the partial transposition for all mixed states which are diagonal in a graph--state basis. We also consider entanglement between different groups of particles and determine the corresponding lifetimes as well as the change of the kind of entanglement with time. This enables us to investigate the behavior of entanglement under re--scaling and in the limit of large number of particles $N \rightarrow \infty$. Finally we investigate the lifetime of encoded quantum superposition states and show that one can define an effective time in the encoded system which can be orders of magnitude smaller than the physical time. This provides an alternative view on quantum error correction and examples of states whose lifetime of entanglement (between groups of particles) in fact increases with the size of the system.

\end{abstract}


\maketitle


\section{Introduction}

Since the early days of quantum mechanics, entanglement has remained at the focus of interest. While entangled states of microscopic samples of matter --such as a few atoms or ions in a trap-- can nowadays be prepared and studied in the laboratory \cite{Fo00}, the question whether entanglement can persist on a macroscopic (i.e. classical) scale is still subject of an ongoing debate. The puzzling consequences of such macroscopic entanglement --highlighted in the notorious gedanken experiment known as ``Schr\"odinger's cat'' \cite{Sr35} by Schr\"odinger in 1935-- and the (as it seems) absence of entanglement in macroscopic objects and hence in our ``classical world'' suggests a mechanism which prevents the persistence of entanglement on a macroscopic scale. It is often argued that decoherence --i.e. interaction of a system with uncontrollable degrees of freedom of some environment-- \cite{Zu03} provides a way to understand the absence of entanglement on a macroscopic scale. In particular, the decoherence rate is believed to grow linearly with the size of the system which would predict a rapid decay of entanglement for systems consisting of many particles.  

Such an argument can easily seen to be valid for certain entangled states, e.g. superposition states of the form
\be
|GHZ\rangle \equiv 1/\sqrt{2}(|0\rangle^{\otimes N}+|1\rangle^{\otimes N}), 
\ee
also called Greenberger--Horne--Zeilinger (GHZ) states which are states of $N$ spins or qubits, that interact with uncontrollable degrees of freedom of the environment, e.g. described by a heatbath. GHZ states can be viewed as simple models of Schr\"odinger cat states and are in fact sometimes called cat--states. For GHZ states, one can indeed show that if $\kappa$ is the decoherence rate of a single qubit, then the rate at which the $N$--qubit state decoheres is given by $\kappa N$. However, the observation that multipartite entanglement becomes more fragile with the size of the system is valid for this specific state only, and a priori it is not clear whether a similar conclusion can be drawn for other multipartite entangled states. 

Moreover, the decoherence rate does not provide complete information about entanglement properties of a system. In the last few years a theory of entanglement has emerged, which allows for a more sophisticated and detailed investigation of the effect of decoherence on the entanglement properties of a multiparticle entangled state. Quite recently, we have shown in Ref. \cite{Du04b} that for GHZ states not only the decoherence rate shows a scaling behavior with the system size, but also the lifetime of distillable entanglement --that is the time after which (distillable) entanglement disappears from a system subjected to decoherence-- in fact decreases with the number of particles $N$, confirming the previous reasoning. On the other hand, we have also shown in Ref. \cite{Du04b} that for a class of other genuine multiparticle entangled states --most notable cluster states \cite{Rau01}--, the lifetime of distillable entanglement does not depend on the number of particles $N$ and thus the size of the system. This is in sharp contrast to the behavior of GHZ states and shows that genuine macroscopic entanglement can indeed persist on timescales which are independent of the size of the system. 

While the investigation in Ref. \cite{Du04b} was limited to a specific decoherence model corresponding in physical terms to individual coupling of particles to a thermal reservoir in the infinite temperature limit, we will show in this article that these observations are largely independent of the specific model of decoherence. In particular, a similar scaling behavior of the lifetime of $N$--party distillable entanglement with the size of the system is obtained for all decoherence models dealing with a coupling of each particle to its own environment (or heat bath), e.g. described by a quantum optical master equation of Lindblad form. The results can even be extended to collective (finite range) couplings of particles to the environment. While for GHZ states we provide analytic results for the lifetime of (distillable) entanglement, we calculate upper and lower bounds on the lifetime for all states which belong to the family of graph states \cite{Rau01,He03}. The lower bound is based on an explicit entanglement distillation protocol, while upper bounds are obtained by three different methods. Using the first method we show that the completely positive map describing the decoherence process becomes entanglement breaking \cite{Shor} after a finite time. This implies that all initially entangled states become separable and thus the lifetime of all kinds of entanglement is finite. The second method is more specific to graph states and shows that graph states suffering from decoherence become separable after a finite time. This is done by using a dynamical description of graph states and by showing that the generating operations become separable. The third method is based on the partial transposition criterion and evaluates when the partial transposition with respect to a certain partition becomes positive. To this aim, we develop a method to calculate the spectrum of the partial transposed operator $\rho^{T_S}$ for any subset $S$ of parties and all density operators $\rho$ which are diagonal in a basis constituted by orthogonal graph states.

We also consider entanglement between $M$ groups of particles, i.e. partitions of the system into M parts. Each of the groups may consist of several particles, which are then considered as a single subsystem with a higher dimensional state space. We analytically determine the lifetime of distillable entanglement between $M$ groups of particles for arbitrary partitionings for GHZ states and again derive lower and upper bounds for all graph states. In this way we study the change of the kind of entanglement with time and e.g. show for GHZ states that the effective size of entanglement, i.e. the maximum number of entangled subsystems, decreases with time and entanglement eventually becomes bipartite before it vanishes completely. If we associate a specific spatial distribution with the particles, e.g. spins distributed on a lattice, one can choose certain partitionings that correspond to a re--scaling of the size of the subsystem, as it is used in statical physics. We study in particular the behavior of distillable entanglement under coarsening of the partitions, that is under re--scaling of the size of the subsystem in the asymptotic limit $N\to \infty$. For cluster states (and all other graph states with constant degree) we show that the lifetime of distillable entanglement is largely independent of $N$ and thus the same on all scales. For GHZ states, however, we find that whenever the size of the subsystems is finite, distillable entanglement vanishes after an arbitrary short time on all scales. Only if the size of the subgroups become macroscopic themselves (in the sense that $N$ systems are divided into a fixed number $M$ of cells whose size $N/M$ grows to infinity as $N\to \infty$) the lifetime of distillable entanglement (between the $M$ cells of macroscopic size) becomes finite and scales to leading order as $1/(\kappa M)$. We also consider the lifetime of encoded entangled states. When considering entanglement properties between groups of particles where each group constitutes a logical qubit of a (concatenated) quantum error correction code, one can define an effective time for the encoded system, that incorporates the error correction procedure. The effective time can be orders of magnitude smaller than the physical time. In this way one can show that the lifetime of entanglement between groups of particles can even increase with the size of the system.

The paper is organized as follows. In Sec.~\ref{decoherence} we introduce decoherence models --most notable individual coupling of a single particle to a reservoir described by a quantum optical master equation of Lindblad form as well as Pauli channels-- which we deal with throughout the paper. In Sec.~\ref{Entanglement} we introduce basic concepts of entanglement theory. In particular, we review the concepts of separability and distillability in multiparticle systems as well as the the partial transposition criterion. We also define lifetime of entanglement with respect to certain partitionings of the system. In Sec.~\ref{GHZ} we determine the lifetime of $N$--party distillable entanglement of GHZ states for decoherence described by depolarizing channels \cite{Du04b} as well as general quantum optical channels. In Sec.~\ref{graph} we first review the concept of graph states \cite{He03} in Sec.~\ref{definitions}, and then derive lower and upper bounds on the lifetime of $N$--party distillable entanglement for graph states subjected to decoherence. We generalize our results to weighted graph states in Sec.~\ref{weighted}. In Sec.~\ref{blockwise} we consider entanglement between groups of particles for GHZ states 
 and determine the lifetime of encoded entangled states in Sec.~\ref{encoded}.
We summarize and conclude in Sec.~\ref{summary}, while 
some technical details e.g. regarding the partial transposition criterion and the corresponding upper bound on the lifetime for mixed states which are diagonal in a graph--state basis can be found in the appendices.

\section{Decoherence models}\label{decoherence}\label{individual}


We consider a single two--level system (qubit) coupled to an environment which is described by a thermal reservoir. The evolution of this qubit is governed by a general quantum optical master equation of Lindblad form
\be
\frac{\partial}{\partial t}\rho = -i[H,\rho] + {\cal L} \rho,
\ee
where $H$ describes the coherent evolution while incoherent processes are described by the superoperator ${\cal L}$. We have
\bea \label{Lindblad}
{\cal L}\rho &=& -\frac{B}{2}(1-s)[\sigma_+\sigma_-\rho + \rho\sigma_+\sigma_- - 2 \sigma_-\rho\sigma_+] \nonumber\\
&& -\frac{B}{2} s [\sigma_-\sigma_+\rho + \rho\sigma_-\sigma_+ - 2 \sigma_+\rho\sigma_-] \nonumber\\
&& -\frac{2C-B}{8} [2\rho -2 \sigma_z\rho\sigma_z], 
\eea
with $\sigma_\pm \equiv 1/2(\sigma_x \pm i \sigma_y)$ and $2C\geq B$. While parameters $B,C$ give the decay rate of inversion and polarization, $s\in [0,1]$ depends on the temperature $T$ of the bath. More precisely $s= \text{lim}_{t\rightarrow \infty} \langle \frac{\mathbf{1}+\sigma_z}{2}\rangle_t$, where $s=1/2$ corresponds to $T=\infty$. It is straightforward to solve this master equation \cite{Br93}. We consider the case $H=0$, i.e. solely decoherence. The eigenoperators and corresponding eigenvalues of ${\cal L}$ can readily be determined and one finds
\bea
{\cal L}\sigma_x&=&-C\sigma_x \\
{\cal L}\sigma_y&=&-C\sigma_y \\
{\cal L}\sigma_z&=&-B\sigma_z \\
{\cal L}\tilde \sigma_0&\equiv& {\cal L}\,\frac{1}{2}[\eins+(2s-1)\sigma_z]= 0.
\eea
For $\rho\equiv \rho(0)=\frac{1}{2}\eins + {\vec a}\cdot \vec\sigma =\tilde \sigma_0 + a_x \sigma_x +a_y \sigma_y + [a_z-(2s-1)/2] \sigma_z$, we have that 
\bea
\rho(t) \equiv e^{{\cal L} t} \rho(0) &=& \tilde \sigma_0 + e^{-Ct} (a_x \sigma_x + a_y \sigma_y) \nonumber \\
&& + e^{-Bt} [a_z -(2s-1)/2] \sigma_z.   
\eea 

Equivalently, one can describe the resulting completely positive map (CPM) ${\cal E}_t$ with $\rho(t)\equiv {\cal E}_t \rho$ as follows:
\bea
{\cal E}_t \rho&=& \sum_{j=0}^3 \lambda_j(t) \sigma_j \rho \sigma_j \nonumber\\
&&+ \mu(t) [\sigma_z\rho \eins + \eins \rho \sigma_z -i \sigma_x \rho \sigma_y + i\sigma_y \rho\sigma_x]
\label{QO}
\eea
with
\bea
\lambda_0(t)&=& \frac{1}{4} (1+2 e^{-Ct} + e^{-Bt}) \\
\lambda_1(t)&=&\lambda_2(t)= \frac{1}{4} (1- e^{-Bt}) \\
\lambda_3(t)&=& \frac{1}{4} (1-2 e^{-Ct} + e^{-Bt}) \\
\mu(t)&=& \frac{2s-1}{4}(1-e^{-Bt}).
\eea
In Sec.~\ref{upper1} we will discuss the entanglement properties of this map and show that it (except for some singular cases) becomes entanglement breaking after some finite time. For $s=1/2$ and $B=C\equiv \kappa$, Eq.(\ref{QO}) describes the coupling of the particle to a thermal bath in the large $T$ limit equivalent to a so--called depolarizing channel (white noise):
\be
{\cal D}\rho= p(t)\rho + \frac{1-p(t)}{4} \sum_{j=0}^3 \sigma_j \rho \sigma_j \hspace{0.2cm} \text{with}  \hspace{0.2cm} p(t) = e^{-\kappa t}\;.
\label{Depol}
\ee
For  $B=0$, $C\equiv \kappa$ and arbitrary $s$, Eq.(\ref{QO}) describes instead the coupling of the particle to a reservoir, that is equivalent to a
 dephasing or phase flip channel:
\be
{\cal D}\rho= p(t)\rho + \frac{1-p(t)}{2}\left( \rho + \sigma_3 \rho \sigma_3 \right)\hspace{0.2cm} \text{with}  \hspace{0.2cm} p(t) = e^{-\kappa t}\;.
\label{Dephas}
\ee
Finally, choosing $s=1$ and $B=2C\equiv \kappa$, one obtains the decay channel (pure damping):
 \be
{\cal D}\rho= E_1 \rho E_1^\dagger \, +\, E_2 \rho E_2^\dagger\; ,
\ee
with the Kraus operators $E_1= \left(\begin{array}{cc} 1 & 0 \\ 0 &\sqrt{1-\gamma} \end{array}\right)$ and
 $E_2= \left(\begin{array}{cc} 0 & \sqrt{\gamma} \\ 0 & 0\end{array}\right)$. Here $\gamma(t)=1-e^{-\kappa t}$ denotes the decay rate for the decay from
 level $|1\rangle$ into level $|0\rangle$.\\
For a system consisting of several particles, we shall be interested in the effect of decoherence on the entanglement properties of this system. We consider as a decoherence model individual coupling of each of the qubits to a thermal bath, where the evolution of the $k^{\rm th}$ qubit is described by the map ${\cal E}_k$ given by Eq.~(\ref{QO}) with Pauli operators $\sigma_j$ acting on qubit $k$. We will be interested in the evolution of a given pure state $|\Psi\rangle$ of $N$ qubits under this decoherence model. That is, the initial state $|\Psi\rangle$ suffers from decoherence and evolves in time to a mixed state $\rho(t)$ given by
\be\label{deoheredState}
\rho(t) \equiv {\cal E}_1 {\cal E}_2 \ldots {\cal E}_N |\Psi\rangle\langle \Psi|.\label{dec}
\ee

 Disregarding a physical description in terms of an underlying interaction between the system and its environment, we will in the following also consider decoherence
 due to individual noise processes of the particles described by some Pauli channel:
 \begin{equation}\label{PauliChannel}
 {\cal D} \rho =  \sum_{i=0}^3 p_i(t)\sigma_i \rho \sigma_i \; \text{with} \; (\sum_{i=0}^3 p_i(t) =1)\, ,
\end{equation}
 These noise channels are of particular interest in quantum information theory, especially in the study of fault-tolerance of quantum computation.
This class contains for example:
\begin{itemize}
\item[1.] for $p_0=\frac{1+3p}{4}$ and $p_1=p_2=p_3=\frac{1-p}{4}$ the above depolarizing channel;
\item[2.] for $p_i=\lambda_i$ the quantum optical channel according to Eq.~(\ref{QO}) with $\mu=0$ ($s=\frac{1}{2}$);
\item[3.] for $p_0=\frac{1+p}{2}$, $p_1=p_2=0$ and $p_3= \frac{1-p}{2}$ the above dephasing channel;
\item[4.] for $p_0=\frac{1+p}{2}$, $p_2=p_3=0$ and $p_1=\frac{1-p}{2} $ the bitflip channel
\end{itemize} 
In the remainder of the paper, we will analyze the time dependence of the entanglement properties of the decohered state $\rho(t)$ for different initial states $|\Psi\rangle$. 
The depolarizing channel is of particular interest, since the decohered state due to an arbitrary noise channel can be further depolarized to some state, which might 
also be obtained by some depolarizing channel. Moreover, among the stated noise models the depolarizing channel is the only channel, that is basis independent,
 i.e. invariant under unitary transformations. 
We will frequently use the Pauli channel and will describe the entanglement properties of $\rho(t)$ in terms of the parameters $p_i$. Nevertheless one has to keep in mind, that the time dependence itself is already included in the parameters $p_i=p_i(t)$.


\section{Separability, distillability and lifetime of $N$--party entanglement} \label{Entanglement}

For the lifetime of entanglement it is not only necessary to specify the underlying decoherence model, but also the very notion 
of multi particle entanglement itself. This is mainly due to the fact that multi party entanglement is a subtle issue in quantum
information theory (see e.g. \cite{multiparty,Du00}). Apart from some special cases, the existence of an entanglement measure, that is satisfying for information theoretic purposes as well as applicable and calculable for mixed states, is still an open problem. 
In the following we will therefore concentrate on the discussion
of two qualitative entanglement criteria. Throughout the paper we will consider $N$ two--level systems (qubits) with corresponding Hilbert space ${\cal H} =(\C^2)^{\otimes N}$. The $N$ particles are distributed among $N$ parties $1,\ldots ,N$. Starting with a pure GHZ or graph state we will consider in Sec.~\ref{GHZ} and \ref{graph} the $N$-party separability and distillability properties of the decohered state $\rho(t)$ (see Eq.~(\ref{deoheredState})):\\ On the one side of the scale the state $\rho(t)$ can still be
 {\it $N$-party distillable entangled}, as it is the case for the corresponding pure states in question. Hereby we call $\rho(t)$  $N$-party distillable, if any
 other true $N$-party entangled state $|\Phi\rangle$ can be obtained (distilled) asymptotically from multiple copies of $\rho$ under local operations and classical communication (LOCC) \cite{Th02,Du00}:
\be\label{Ndistillable}
\rho^{\otimes k} \;\longrightarrow_{\text{LOCC}} \; |\Phi\rangle\langle \Phi|\; .
\ee
We remark that in the multi--copy case all true $N$--party entangled states are equivalent since they can be transformed into each other by LOCC. That is, the condition that any true $N$ party entangled state can be created can be replaced by the condition that some $N$--party entangled state, e.g. the initial pure state, can be created. Disregarding the practicability of the underlying distillation protocol, the state $\rho(t)$ is then as useful as any other entangled state and therefore can in principle be regarded as a universal resource for quantum information processing such as quantum communication. 

On the other end of the scale, $\rho(t)$ might have also become completely separable or classical in the sense that it can be described by a classical mixture of product states, i.e. $\rho$ is  {\it $N$-party separable}, if
\be\label{Nseparable}
 \rho (t) = \sum_k\, p_k\, \rho_k^{(1)}\otimes \rho_k^{(2)}\otimes \ldots \otimes \rho_k^{(N)} \; .
\ee
If a state is completely separable, it is no longer entangled whatsoever. In between these two extremal cases, $\rho(t)$ can contain different types of {\it blockwise entanglement}, which we will discuss in more detail in Sec.~\ref{blockwise}. 
There we will consider different partitionings of particles into $M$ groups ($M\leq N$), where each group forms a subsystem with a higher dimensional state space and consists of several particles. {\it $M$-party distillability [separability]} can then be defined
 {\it with respect to a given partitioning} in a similar way, where the notion of {\em local} operation has to be adapted accordingly. Moreover we will call $\rho(t)$ {\it $M$-party distillable}, if there exists at least one partitioning, with respect to which $\rho(t)$ is $M$-party distillable. 

Based on the notion of $M$--party separability and distillability, one can define lifetime of entanglement. A $N$--party state $|\Psi\rangle\langle\Psi|$ which is subjected to decoherence for time $t$ evolves into a mixed state $\rho(t)$. The lifetime of $N$--party distillable entanglement is given by the time after which the state $\rho(t)$ becomes non--$N$--party distillable. This implies that lower bounds on the lifetime of distillable entanglement can be obtained by showing that the state $\rho(t)$ is distillable, while an upper bound can be obtained by proving non--distillability of $\rho(t)$. When considering partitions of the system into $M$ groups, the lifetime of $M$--party entanglement with respect to a given partition is defined accordingly. We refer to the lifetime of $M$--party entanglement as the time after which $\rho(t)$ is non--distillable with respect to {\em all} $M$--party partitions. In a similar way, one can define a lifetime with respect to the separability properties of $\rho(t)$.

In order to determine entanglement properties of the mixed states in question, we will continuously make use of the partial transposition criterion \cite{Pe96,Ho97}, an entanglement criterion which provides necessary conditions for distillability and separability. The partial transposition is defined for bipartite systems only, while a system can in general consist of several parties. Making use of the concept of partitionings of the system, in particular considering all bipartitionings, one can use the partial transposition criteria also for multipartite states. Let $A$ denote a subset of $m$ parties $k_1, \ldots ,k_m$. In general, given an operator $X$ acting on $\C^{d_A}\otimes\C^{d_B}$, we define the partial transpose of $X$ with 
respect to the first subsystem in the basis 
$\{|1\rangle,|2\rangle,\ldots,|d_A\rangle\}$, $X^{T_A}$, as follows:
\be
X^{T_A} \equiv \sum_{i,j=1}^{d_A}\sum_{k,l=1}^{d_B}
\langle i,k|X|j,l\rangle \; |j,k\rangle\langle i,l|.
\ee
A hermitian operator $X$ has a non--positive [positive] partial transpose 
(NPT) [(PPT)] if $X^{T_A}$ is not positive [positive] respectively. That is, $X^{T_A}$ is NPT if there exist some 
$|\Psi\rangle$ such that $\langle\Psi|X^{T_A}|\Psi\rangle <0$. 

The positivity of the operator $\rho^{T_A}$ gives a necessary criterion for separability, whereas the non-positivity of $\rho^{T_A}$ is necessary for the
distillability of the density operator $\rho$. In particular, if a bipartite density operator is PPT, then it is certainly not  distillable \cite{Ho97}. This implies \cite{Du00} that if a multiparticle density operator $\rho$ is PPT with respect to at least one bipartite partition, then $\rho$ is certainly not $N$--party distillable. On the other hand, positivity of all bipartite partitions is a necessary condition for $N$--party separability.  In the case of two dimensional systems $\mathbb{C}^2\otimes\mathbb{C}^2$ the PPT [NPT] criterion is necessary {{\it and} sufficient for separability [distillability] \cite{Pe96,Ho96}. A detailed discussion of the application of the partial transposition criteria to multipartite systems can be found in Ref. \cite{Du00}.

\section{Lifetime of N-party entanglement in GHZ states}\label{GHZ}

We start by considering the lifetime of $N$--qubit GHZ states
\be
|GHZ\rangle=1/\sqrt{2}(|0\rangle^{\otimes N} + |1\rangle^{\otimes N}).\label{GHZstate}
\ee
These states are special examples of states, that maximally violate multi-partite Bell inequalities \cite{BI}.
GHZ states have also become an interesting resource for multi-party quantum communication, e.g. in the context of secret sharing and secure function evaluation
 \cite{Secure}. Moreover they can be used to improve frequency standards \cite{Frequency}.
For the class of $N$-party GHZ states the lifetime of most of the above entanglement properties can be determined analytically.

\subsection{Large $T$ limit of reservoir}\label{largeT}

We start by reviewing the results of \cite{Du04b} and consider a model of decoherence with individual coupling of each of the particles to a thermal reservoir in the large $T$ limit. The process for a single qubit is described by Eq.~(\ref{Depol}) and corresponds to white noise with time dependent parameter $p\equiv p(t)=e^{-\kappa t}$ where $\kappa$ is a coupling constant. That is, we consider the state $\rho(t)$ given by Eq.~(\ref{dec}) where the CPM ${\cal E}_k$ is given by the the depolarizing channel ${\cal D}_k$ (Eq.~(\ref{Depol})).   
It is straightforward to evaluate the effect of decoherence on this kind of states \cite{Si02,Du00}. One finds that $|GHZ\rangle$ evolves to a state $\rho(t)$ 
\be
\rho(t)&=&\sum_{ k_i=0}^1\lambda_{k_1\ldots k_{N}} P_{k_1\ldots k_{N}} + \mu (\sigma_+^{\otimes N} + \sigma_-^{\otimes N}),
\label{rhot}
\ee
with 
\be
P_{k_1k_2\ldots k_{N}}&=&|k_1k_2\ldots k_{N}\rangle\langle k_1k_2\ldots k_{N}|, \ee
and $\sigma_\pm = (\sigma_x \pm i\sigma_y)/2$, i.e. $\sigma_+^{\otimes N}= |00\ldots 0\rangle\langle 11\ldots 1|$. It turns out that the coefficients $\lambda_{k_1k_2\ldots k_{N}}$ fulfill $\lambda_{k_1k_2\ldots k_{N}}=\lambda_{\bar k_1\bar k_2\ldots \bar k_{N}}$, where $\bar k_j =1-k_j$. In addition, $\lambda_{k_1k_2\ldots k_{N}}$ depend only on $k\equiv\sum_{j=1}^{N} k_j$, that is $\lambda_{k_1k_2\ldots k_N} = \lambda_{l_1l_2\ldots l_N} \equiv \lambda_k$ if $\sum_{j_1}^{N} k_j=\sum_{j_1}^{N} l_j =k$. This implies that $\lambda_k = \lambda_{N-k}$ and one finds
\bea
\lambda_k & = & \frac{1}{2^{N+1}}\left((1+p)^k(1-p)^{N-k}+(1+p)^{N-k}(1-p)^k\right), \cr
\mu &=& \frac{p^N}{2}.
\label{lambdakGHZ}
\eea
States of the form Eq.~(\ref{rhot}) with $\lambda_{k_1\ldots k_{N}} = \lambda_{\bar k_1\ldots \bar k_{N}}$ can equivalently be written as density operators which are diagonal in a basis consisting of orthogonal GHZ states. Such density operators have been completely characterized with respect to their entanglement properties in Ref.~\cite{Du00}. In particular, it was shown that these states are $N$--party distillable [separable] if and only if the partial transpose with respect to all possible partitions is non--positive [positive], respectively. One readily finds that the partial transposition with respect to a group $B_k$ which contains exactly $k$ parties is positive, $\rho(t)^{T_{B_{k}}} \geq 0$,
if and only if \cite{Du00} $\mu^2 \leq \lambda_k \lambda_{N-k}$, i.e.
\be 
p^N \leq 2 \lambda_{k}.\label{cond}
\ee
Making use of the fact that 
\be
\lambda_1 \geq \lambda_2 \geq \ldots \geq \lambda_{[N/2]},\label{condlambda}
\ee
it is now straightforward to determine the lifetime of distillable $N$--party entanglement as well as the time when the state becomes fully separable. From Eqs.~(\ref{cond}, \ref{condlambda}) follows that the lifetime of distillable $N$--party entanglement is determined by Eq.~(\ref{cond}) with $k=1$, as the partial transpose with respect to the partition one party - $(N-1)$ parties is the first one to become positive. Similarly, Eq.~(\ref{cond}) with $k=[N/2]$ determines the time after which the state $\rho(t)$ becomes fully separable, as the partial transposition with respect to the partition $N/2-N/2$ parties is the last one to become positive.

One observes that the critical value $p_{\rm crit}\equiv e^{-\kappa t_{\rm crit}}$, at which the partial transposition with respect to one party becomes positive increases with $N$.
This implies that for $t\geq t_{\rm crit}\equiv \tau$ the state is no longer $N$--party distillable entangled and thus the lifetime $\tau$ of true $N$--party entanglement decreases with the size of the system as expected (see Fig.~\ref{fig:upperGHZ1} and \ref{fig:upperGHZ2}). Note that finding the threshold value $p_{\rm crit}$ for a given $N$ exactly is equivalent to finding the roots of a polynomial of degree $N$ (which can be done efficiently numerically). One can obtain analytic upper and lower bounds on $p_{\rm crit}$ by approximating $\lambda_k$ by 
$(1+p)^{N-k}(1-p)^k]/2^{N+1}$ or $2(1+p)^{N-k}(1-p)^k]/2^{N+1}$ respectively, which is done explicitly in Sec.~\ref{block}.


\begin{figure}[ht]

\includegraphics[width=8cm]{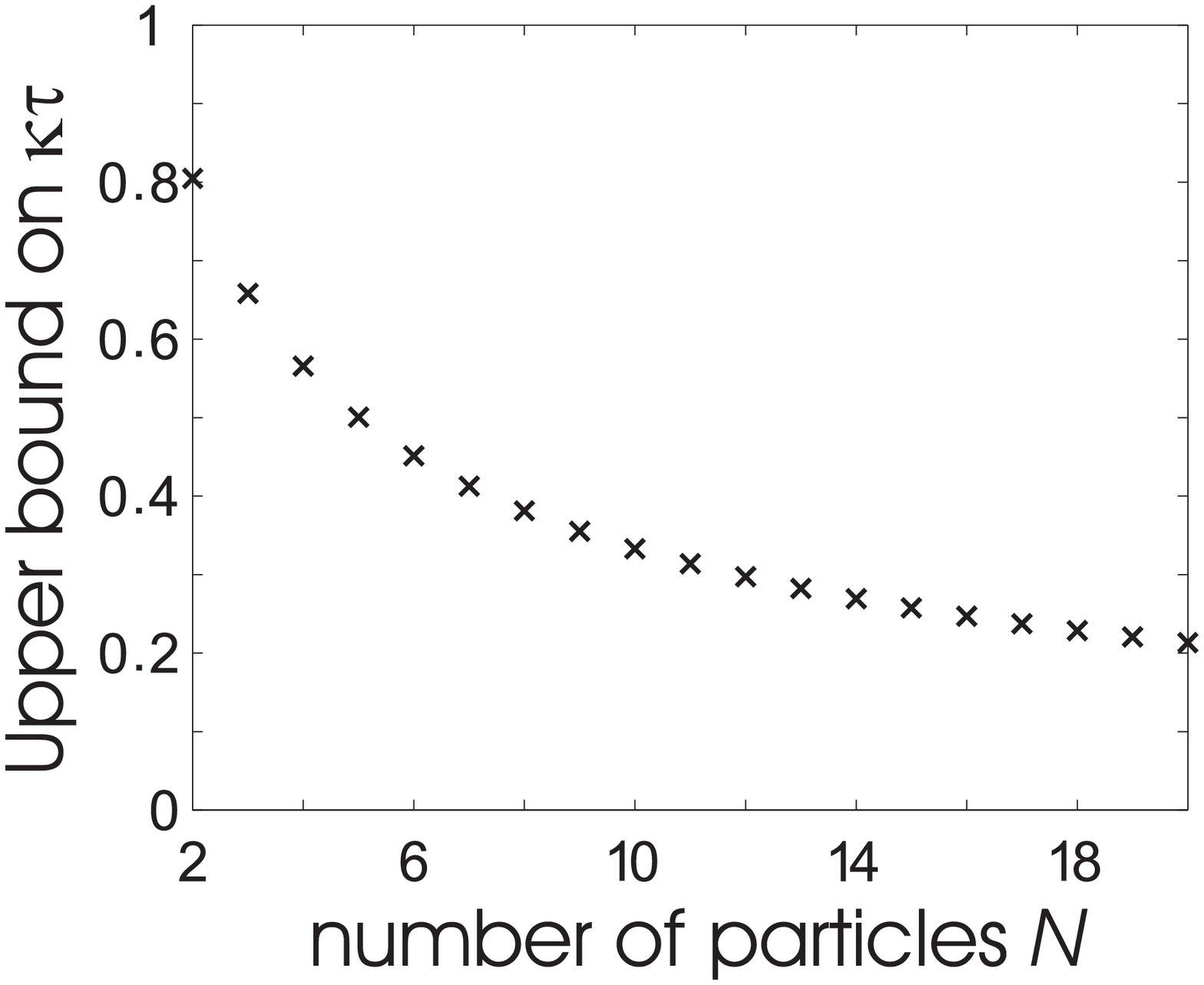} \\ 
\caption{\label{fig:upperGHZ1} Upper bound on lifetime $\kappa \tau$ of $N$--party entanglement.}
\includegraphics[width=8cm]{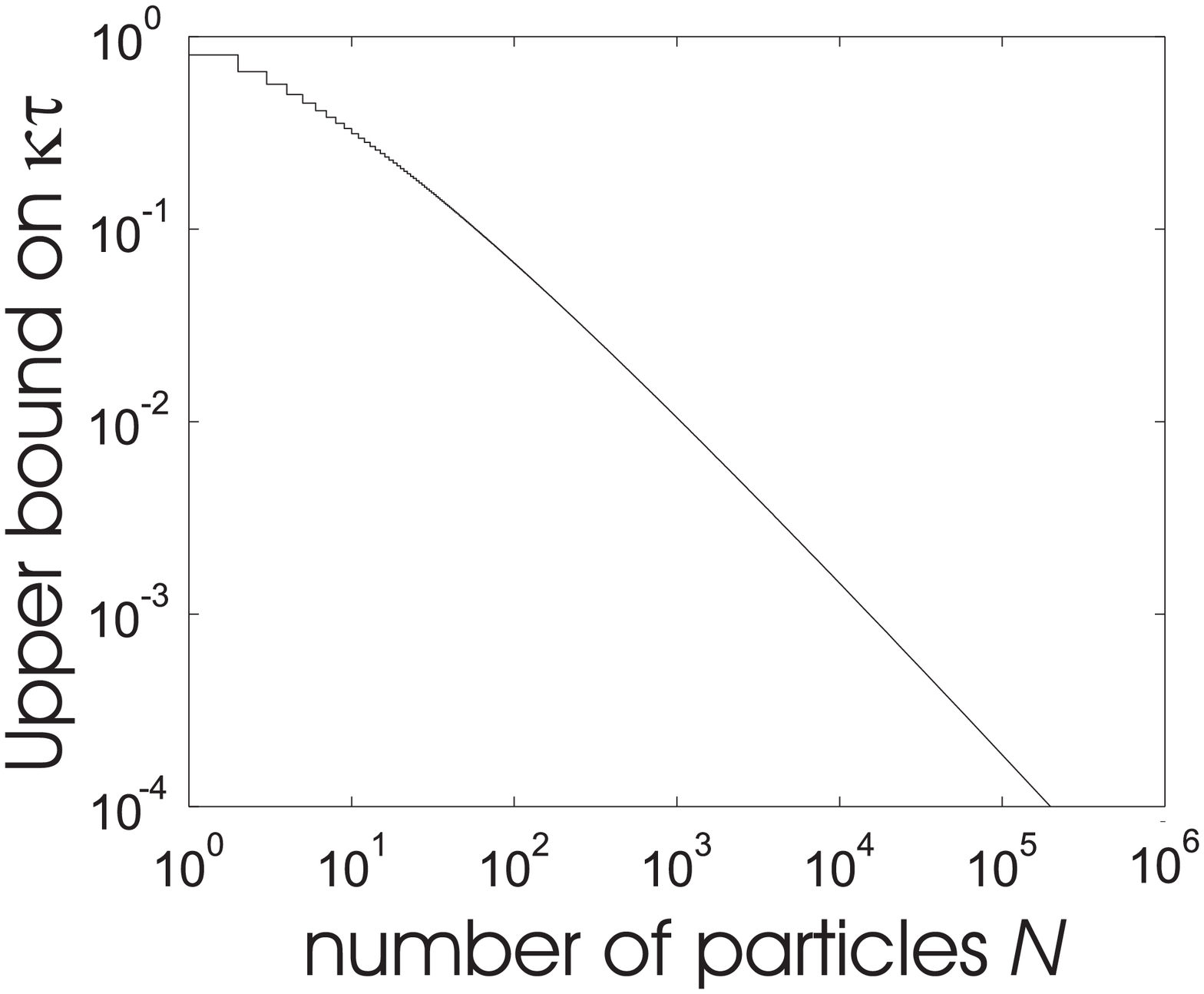} 
\caption{\label{fig:upperGHZ2}Same as Fig. 1 but with double--logarithmic axis.\\ Note that the same figures also display an upper bound on the lifetime $\kappa \tau$ of $M$--party entanglement in systems with $N \rightarrow \infty$ particles for different $M$, as discussed in Sec.~\ref{blockwise}. In this case the numbers on the 
$x$-axis have to be considered as the number $M$ of $M$-party entanglement in question.
}
\end{figure}

\subsection{Arbitrary individual coupling to the environment}\label{GHZ_QO}
We will now investigate the lifetime of distillable entanglement for more general decoherence models. While we continue to assume an individual coupling of particles to independent environments --an assumption which is particularly well fulfilled if the entangled states in question are distributed among several parties--, we consider now couplings which are described by arbitrary quantum optical master equations of Lindblad form (see Sec.~\ref{individual}). These models include as particular instances decay channels, phase flip channels and depolarizing channels.
We consider the influence of decoherence --described by the CPM  Eq.~(\ref{QO})-- on the GHZ state of $N$ particles, i.e. the entanglement properties of the density operator $\rho(t)$ which is given by Eq.~(\ref{dec}). 
We use that one can write $\rho(0)\equiv |GHZ\rangle\langle GHZ|$ as
\be
\rho(0)=\frac{1}{2}(P_0^{\otimes N}+P_1^{\otimes N} + \sigma_+^{\otimes N} + \sigma_-^{\otimes N}),
\ee
where $P_0 =|0\rangle\langle 0| = \frac{\mathbf{1} + \sigma_z}{2}, P_1 =|1\rangle\langle 1|= \frac{\mathbf{1} - \sigma_z}{2}$ and $\sigma_\pm =(\sigma_x \pm i\sigma_y)/2$. It is not difficult to see that the action of the map ${\cal E}$ (Eq.~(\ref{QO})) is given by
\bea
{\cal E}P_0 &=& aP_0+(1-a)P_1, \nonumber\\
{\cal E}P_1 &=& (1-c)P_0+cP_1, \nonumber\\
{\cal E}\sigma_\pm &=& b \sigma_\pm,
\eea
where we introduced the new variables $a,b,c$ which are given by
\bea \label{abc}
a&=&s+(1-s)e^{-Bt},\nonumber\\
b&=&e^{-Ct},\nonumber\\
c&=&(1-s)+se^{-Bt}.
\eea
It is now straightforward to determine the action of the map ${\cal E}_1{\cal E}_2\ldots {\cal E}_N$ on the state $\rho(0)$. One finds that the resulting density operator is of the form Eq.~(\ref{rhot}). The coefficients $\lambda_{k_1 k_2\ldots k_{N}}$ only depend on $\sum_{j=1}^{N} k_j$, where
\bea
\lambda_k&=&\frac{1}{2}\left(c^k(1-c)^{N-k}+ (1-a)^ka^{N-k}\right)\nonumber\\
\mu&=&\frac{b^N}{2}
\eea
The condition that the partial transposition with respect to $k$ parties is positive, $\rho^{T_{A_k}} \geq 0$ reads 
\be
[b^N/2]^2 \leq \lambda_k \lambda_{N-k}.\label{condPPTg}
\ee
We remark that in contrast to the discussion in Sec.~\ref{largeT}, here we have $\lambda_k \not=\lambda_{N-k}$. This means that non--positive [positive] partial transposition with respect to all partitions is no longer a sufficient condition for $N$--party distillability [separability] respectively. However, one can still use the partial transposition criterion to obtain lower and upper bounds on the lifetime of distillable entanglement. In particular, if the partial transposition with respect to at least one partition is positive, then the state $\rho(t)$ is certainly no longer $N$--party distillable. 

To obtain an upper bound on the lifetime of GHZ states, we make use of the following facts:  (i) $\lambda_k \lambda_{N-k} \geq (ac)^{N-k}[(1-a)(1-c)]^k/4$ and (ii) $\exp(-N Bt)/4 \geq [b^N/2]^2$. While (i) can be checked by direct computation, (ii) follows from $2C -B \geq 0$ (see Sec.~\ref{individual}). Using (i) and (ii) together with Eq.~(\ref{condPPTg}), one obtains that $\rho(t)$ certainly has positive partial transposition with respect to any group of $k$ parties if 
\be
N \geq k \frac{\log(ac)-\log[(1-a)(1-c)]}{\log(ac)+B t},\label{condN}
\ee
provided that $s\not=0,1$ and $B>0$. We remark that the (singular) case $s=0$ corresponds to a decay channel, and for such a channel we have that the state $\rho(t)$ has non--positive partial transposition for all times $t$. Whenever the temperature of the bath is however not zero (i.e. $s\not=0,1$) we have that for any time $t$ there exists a finite number $N_0$ (given by the right hand side of Eq.~(\ref{condN}) with $k=1$) such that for $N \geq N_0$ particles the state $\rho(t)$ is certainly no longer distillable. Thus we have --as in the case of depolarizing channels-- a scaling of the (upper bound on) lifetime of distillable entanglement with the number of particles $N$. If $N$ is sufficiently large, the (upper bound) on the lifetime goes to zero. 

A lower bound on the lifetime of $\rho(t)$ can be obtained as follows: We have that a state of the form Eq.~(\ref{rhot}) can be  depolarized by means of a (stochastic) sequence of local operations and classical communication (see Ref. \cite{Du00}) such that the resulting state has new coefficients $\tilde \lambda_{k_1k_2 \ldots k_N}$ which fulfill
\be
\tilde \lambda_{k_1k_2 \ldots k_N} = \frac{\lambda_{k_1k_2 \ldots k_N} + \lambda_{\bar k_1 \bar k_2 \ldots \bar k_N}}{2}, 
\ee
and hence $\tilde \lambda_k=\tilde \lambda_{N-k}$. It follows that the depolarized state $\tilde \rho(t)$ is distillable if 
\be
[b^N/2] > \tilde \lambda_k,
\ee 
for all $k$. One can upper bound $\tilde \lambda_k$ by $\tilde \lambda_k \leq \lambda'_k \equiv \max(a^k(1-a)^{N-k}, a^{N-k}(1-a)^k, c^k(1-c)^{N-k}, c^{N-k}(1-c)^k)$ and obtains that $\rho(t)$ is distillable if $b^N > 2\lambda_k'$. By taking the logarithm of this equation, one obtains a bound on the number of particles $N$ such that the state remains distillable for a time $t$.

\section{Lifetime of N-party entanglement in graph states}\label{graph}
 
In the previous section the class of generalized GHZ states was shown to have a lifetime of entanglement, that decreases (except in some singular cases) with the number of particles $N$ in the system. 
We will now discuss the lifetime of $N$-party entanglement in graph states and show, that for a significant subclass such as the cluster states the lifetime of distillable
 entanglement is essentially independent of $N$. After recalling some basic definitions and notations, we will first derive a lower bound to $N$-party distillable entanglement by providing an explicit distillation protocol. We will then use three different techniques to establish upper bounds to the lifetime of $N$-party entanglement. These methods apply to different decoherence processes and are interesting in their own, since they might find applications also in other problems not directly related to lifetime of states under decoherence. Finally we will extend our results to a more general class of so called weighted graph states. 

\subsection{Basic definitions and examples}\label{definitions}

Graph states are multi-particle spin states of distributed quantum systems with interesting applications in quantum information theory:
 Special instances of graph states are codewords of quantum error correcting codes, which protect quantum states against decoherence in quantum computation. Up to local unitaries all stabilizer states can be represented as graph states~\cite{graphCodes}. For example, the CSS--codes correspond to the class of so called 2-colorable graphs~\cite{Lo04}. For this class of graph states entanglement purification procedures are known~\cite{Du04b}. These protocols even work in the case of noisy local control operations. Finally the class of cluster states are known to be a universal resource for quantum computation in the one-way quantum computer \cite{oneWay}.
For the study of genuine multi-partite entanglement graph states are particularly useful, since they allow for an efficient description even in the regime
of many parties:
Thereby the graph essentially encodes an interaction pattern between the particles. Let $G=(V,E)$ be a graph, which is a set of $N$ vertices $k\in V$ connected by edges $\{k,l\}\in E$ that specify the neighborhood relation between the vertices. Starting from the state $|+ \rangle^{V} := \bigotimes_{k \in V} |+\rangle^{(k)}$, where $|+\rangle=\frac{1}{\sqrt{2}}\left(|0\rangle + |1\rangle \right)$ denotes the eigenstate of $\sigma_x$ with eigenvalue $+1$, the graph state $|G\rangle$ is obtained by applying a sequence of Ising-type interactions 
\be \label{Ising}
 U_{kl}\equiv e^{-i\frac{\pi}{4} \left(\mathbf{1}^{(k)} - \sigma_z^{(k)}\right)\otimes \left(\mathbf{1}^{(l)} - \sigma_z^{(l)}\right)}\ee  according to the interaction pattern specified by the graph, i.e.
\be |G\rangle = \prod_{\{k,l\}\in E} U_{kl}\, |+ \rangle^{ V} \ee
 Graph states occur e.g. as a result of the Ising interaction between 
neighboring spins on a lattice after a specific interaction time \cite{Rau01}. An example for a realization of such a system is based on neutral atoms in optical lattices 
\cite{Realisation}. Alternatively graph states can be specified in terms of their stabilizer: For this let $N_k =\left\{l \in V\, |\, \{k,l\} \in E \right\}$
 denote the set of neighbors of $k$.
Then the graph state $|G\rangle$ is the unique state in $({\mathbb{C}}^2)^{\otimes V}$, that is the common eigenstate to the set of independent
 commuting observables:
\begin{equation}\label{stab}
K^G_{k} \equiv \sigma_x^{(k)}\prod_{l\in N_k}\sigma_z^{(l)},
\end{equation}
where the eigenvalues to all $k \in V$ are $1$. The stabilizer $\mathcal{S}_G$ of the state is thus generated by the set $\{K^G_k\, |\, k\in V\}$, which implies \be |G\rangle\langle G| = \sum_{\sigma \in \mathcal{S}_G} \sigma \; .\ee
In order to obtain a complete basis for $({\mathbb{C}}^2)^{\otimes V}$ we will also consider the eigenstates $|U\rangle_G = \sigma_z^U |G\rangle$ of $K^G_k$ according to different eigenvalues $U_k$, i.e. \be K^G_k |U\rangle_G = (-1)^{U_k} \, |U\rangle_G  \; . \ee

Here and in the following, sets $U\subseteq V$ as an upper index for operators will label those vertices where the operator acts non-trivially, e.g. 
\be
\sigma_z^U=\bigotimes_{k \in U} \sigma_z^{(k)}\; .
\ee
Moreover we will denote sets $U$ and their corresponding binary vectors $U=(U_k)_{k\in V} \simeq (U_1,\ldots, U_{N})$ over $\mathbb{F}_2^V$(the integer field modulo $2$) with the same symbol. Finally $k$ will also denote both the vertex and the corresponding one-element set $\{k\}$. 
In this notation the stabilizer generators can be written as $K^G_k =\sigma_x^k \sigma_z^{N_k}$ and the original graph state is just that with an error syndrome corresponding to the empty set $0$, i.e. $|G\rangle=|0\rangle_G$. This is also notationally advantageous, since we will use both set and binary operations:
E.g. for $A,B \in \mathcal{P}(V) \cong \mathcal{F}_2^V$ we will write $A\cup B$, $A\cap B$ and $A\setminus B$ ($\bar{A}\equiv V\setminus A$) for the union, intersection and difference (complement) as well as $A+B$ and $\langle A, B\rangle$ for the addition and the scalar product modulo $2$. 
The neighborhood relation in a graph is also often represented in terms of its adjacency matrix $\Gamma$:
\begin{equation}
    \Gamma_{kl} =
    \left\{
    \begin{array}{ll}
    1,& \text{ if $\{k,l\}\in E$,}\\
    0 & \text{otherwise}.
    \end{array}
    \right.
\end{equation}
In the spirit of the above notation we can therefore also write $
    K^G_{k} =\sigma_x^{(k)}\prod_{l\in V}
(\sigma_z^{(l)})^{\Gamma_{kl}}=\sigma_x^k\sigma_z^{N_k}=\sigma_x^k\sigma_z^{\Gamma\, k}
$.

Coming to some examples, we first note, that the class of multi-party GHZ states in Sec. \ref{GHZ} is contained in the class of graph states, since the GHZ state
 in Eq.(\ref{GHZstate}) can be transformed by local unitaries into graph states corresponding to the graphs depicted in Fig.~\ref{fig:GHZ}.
\begin{figure}[th]
\includegraphics[width=3cm]{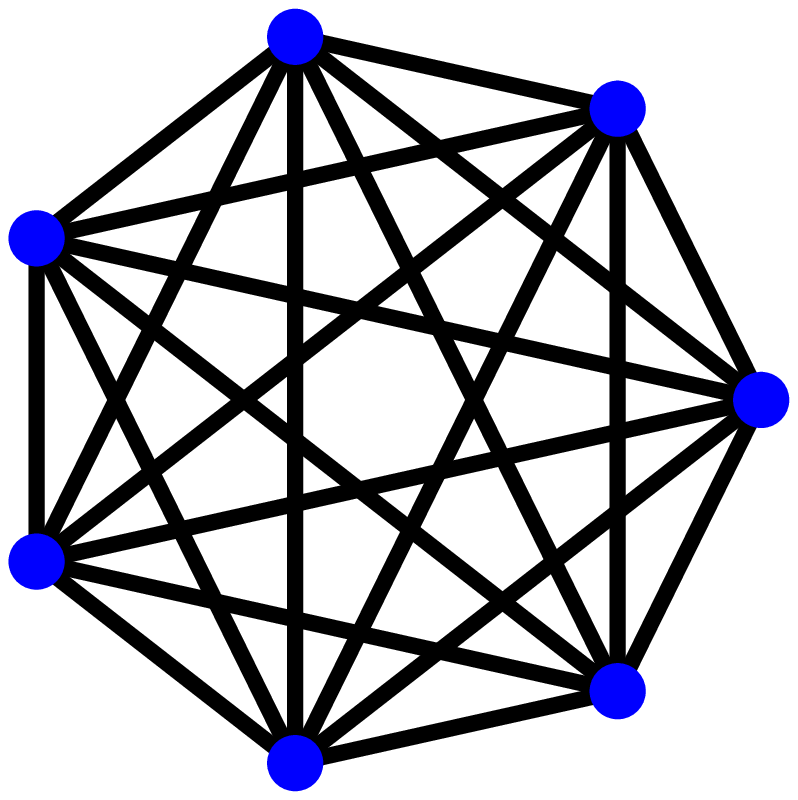}
\hspace{1cm} 
\includegraphics[width=3cm]{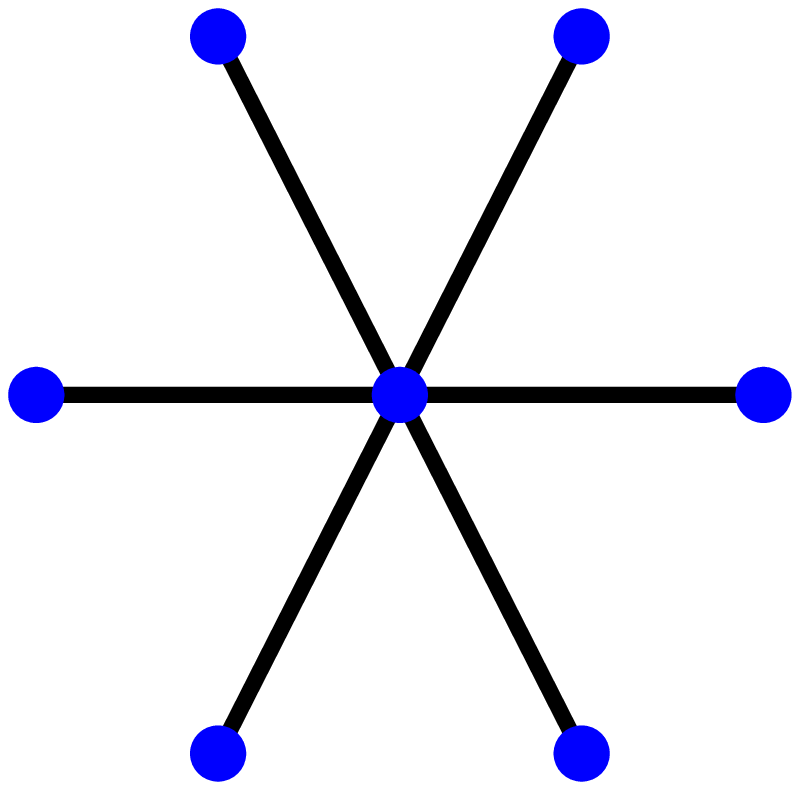}  
\caption{\label{fig:GHZ} The graph states corresponding to the complete graph and the star graph are equivalent to the GHZ state in Eq.~(\ref{GHZstate}) up to some local unitaries \cite{He03}.}
\end{figure}
 When considering decoherence of a locally equivalent state, we remark that the underlying noise process has to be adapted according to the local unitary transformation. From this point of view the depolarizing channel in Eq.~(\ref{Depol}) has the advantage that it is invariant under local unitary transformation and hence is basis independent.
 In the following we will also consider the class of cluster states in $1$, $2$ or $3$ dimensions (see Fig.~\ref{fig:Cluster}), which are of particular interest in the
 context of 'one-way' quantum computation \cite{oneWay}. For more examples and a discussion of equivalence classes of graph states under local unitaries and/or graph isomorphies we refer to \cite{He03,Ma03}.  
\begin{figure}[th]
\begin{picture}(230,100)
\setlength{\unitlength}{1cm}
\put(0,0.2){\includegraphics[width=3cm]{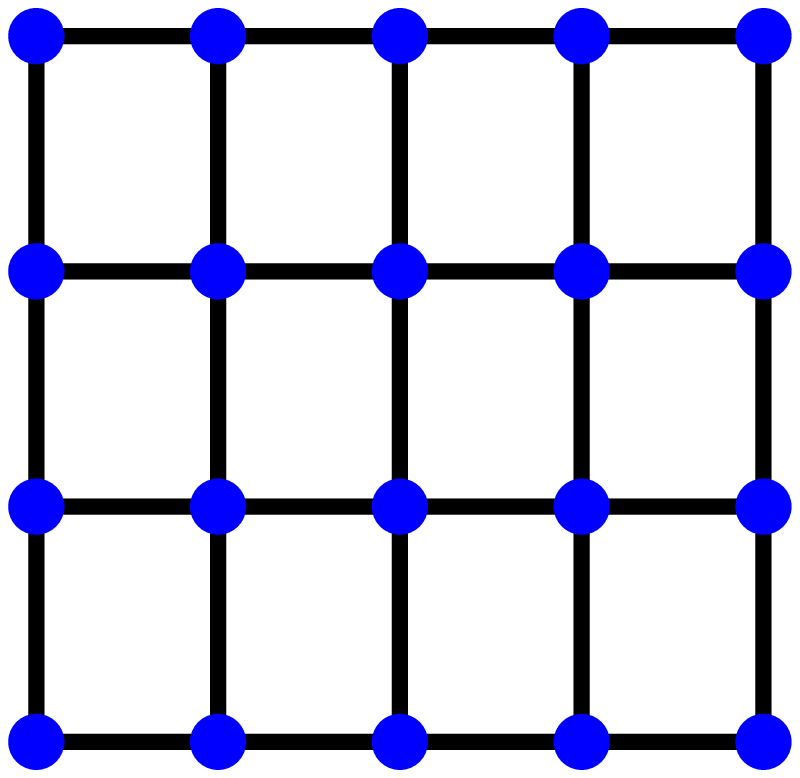}} 
\put(4,0){\includegraphics[width=4cm]{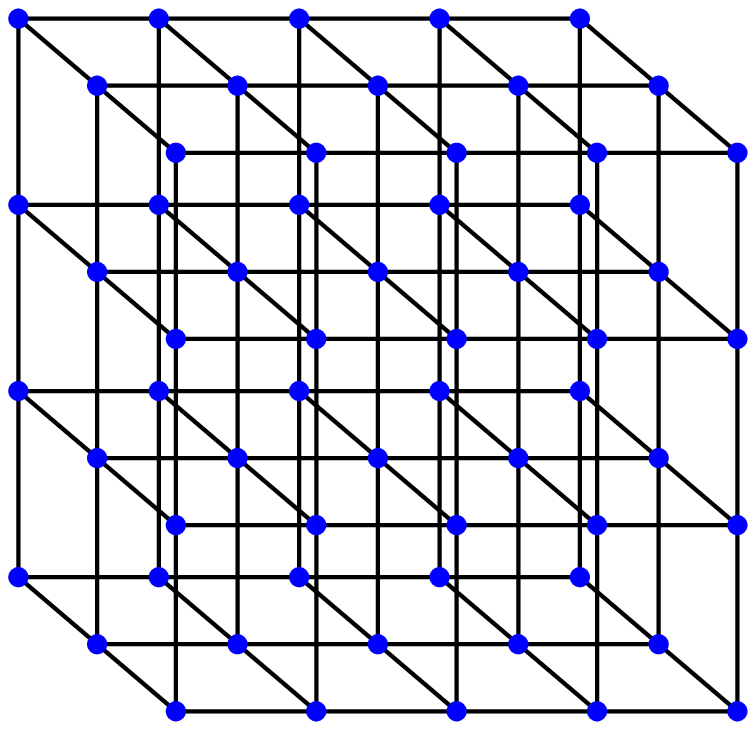}}  
\end{picture}
\caption{\label{fig:Cluster} Cluster states in $2$ and $3$ dimensions form a universal resource for quantum computation in the framework of the one-way quantum computer
 \cite{oneWay}.}
\end{figure}
We will now discuss the entanglement properties from Sec.~\ref{Entanglement} for the state $\rho(t)$ (see Eq.~(\ref{deoheredState})), that is produced
by different decoherence processes described in Sec.~\ref{individual} acting on different types of graph states.

\subsection{Lower Bound: An explicit distillation protocol}\label{lower}

We establish a lower bound on the lifetime of graph states by considering an explicit distillation protocol. In order to show that a mixed state $\rho(t)$ is $N$--party distillable, it is sufficient to show that maximally entangled pairs shared between any pair of neighboring parties can be distilled. This is due to the fact that these pairs can be combined by means of local operations (e.g. by teleportation) to create an arbitrary $N$--party entangled pure state. We emphasize that we use the distillation of neighboring pairs only as a tool to show $N$--party distillability. This does, however, not imply that the entanglement contained in the cluster state was in some sense only ``bipartite''. One could in principle also use direct multi--party entanglement purification protocols, e.g. the one introduced in Ref.~\cite{Du04}, however the conditions under which these protocols are applicable are in general more complicated to determine.

First we will consider the case of decoherence of the particles due to the same individual Pauli channel $ {\cal D} \rho =  \sum_{i=0}^3 p_i(t)\sigma_i \rho \sigma_i $ and
we  will show then how to extend these results to more general decoherence models. We will essentially follow the ideas used in \cite{Du04b}, in which the corresponding result was shown for the case of a depolarizing channel, and make use of the following facts:

{\bf (i)} Measuring all but two neighboring particles, say $k,l$ of a graph state $|G\rangle$ in the eigenbasis of $\sigma_z$ results in the creation of another graph state with only a single edge $\{k,l\}$ \cite{He03}. That is, the resulting state of particles $k,l$ is up to local $\sigma_z$ operations equivalent to a maximally entangled state of the form 
\be\label{Phi} |\Phi\rangle \equiv \frac{1}{\sqrt{2}}\left(|0\rangle_x|0\rangle_z + |1\rangle_x|1\rangle_z\right) \; ,\ee
 where $|i\rangle_x$ [$|i\rangle_z$] denote eigenstates of $\sigma_x$ [$\sigma_z$] respectively. 

{\bf (ii)} The action of a Pauli channel ${\cal D}_k$ acting on particle $k$ of a graph state can equivalently be described by a map ${\cal M}_k$ whose Kraus operators only contain products of Pauli matrices $\sigma_z$ and the identity, where here $\sigma_z$ may act on particle $k$ and its neighbors, i.e. particles which are (in the corresponding graph) connected by edges to particle $k$. 

Observation {\bf (ii)} follows from the fact that $\sigma_x^{j}|U\rangle_G = (-1)^{U_j}\sigma_x^{j} K^G_j |U\rangle_G$, where $\sigma_x^{j} K^G_j$ is an operator which contains only products of $\sigma_z$ operators at neighboring particles of particle $j$, and the identity otherwise. Similarly, the action of $\sigma_y^{j}$ on graph states is up to a phase factor equivalent to the action of an operator which contains only products of $\sigma_z$ operators acting on particle $j$ and all its neighbors. 
That is,
\bea
\label{Mk}
{\cal D}_k|U\rangle_G\langle U| &=&  \sum_{j=0}^3 \,p_j(t)\, \sigma_j^{k} |U\rangle_G\langle U| \sigma_j^{k} \nonumber\\
={\cal M}_k |U\rangle_G\langle U|&=&  \sum_{j=0}^3 \,p_j(t)\, S_j^{k} |U\rangle_G\langle U|  S_j^{k},
\eea
with 
$
S_0^{k}=\sigma_0^{k},
 S_1^{k}=\sigma_3^{N_k}, 
 S_2^{k}= \sigma_3^{N_k \cup k}$ and $ 
 S_3^{k}=\sigma_3^{k}.
$
\begin{figure}[th]
\includegraphics[width=3cm]{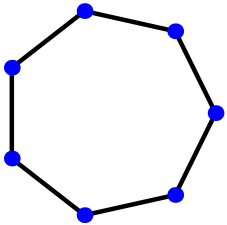}  
\caption{\label{fig:Ring} Ring with seven qubits.}
\end{figure}
In the case of a ring (Fig.~\ref{fig:Ring}), for example, we  have $S_0^{k}=\eins, S_1^{k}=\sigma_3^{(k-1)}\sigma_3^{(k+1)}, S_2^{k}=\sigma_3^{(k-1)}\sigma_3^{(k)}\sigma_3^{(k+1)}$ and $ S_3^{k}=\sigma_3^{(k)}$.

We now apply ${\bf (i),(ii)}$ to establish a sufficient condition when bipartite entanglement between neighboring particles can be distilled from the state 
\be \rho(t) = \mathcal{M}_1 \mathcal{M}_2 \ldots \mathcal{M}_N \, | G \rangle \langle G|\; .\ee
This allows us to obtain a lower bound on the lifetime of graph states. We concentrate on two specific neighboring particles, say $k$ and $l$. One performs measurements in the eigenbasis of $\sigma_z$ on all but particles $k$ and $l$ (We remark that measurements on all neighboring particles of particles $k,l$ would also be sufficient). It follows from {\bf (i)} and {\bf (ii)} that these measurements commute with the action of the CPM ${\cal M}_1{\cal M}_2\ldots {\cal M}_N$ on the graph state (which equivalently describes the action of Pauli channels on these states). That is, the resulting state after the measurements is given by ${\cal M}_1{\cal M}_2\ldots{\cal M}_N |\tilde\Phi\rangle_{k,l}\langle\tilde\Phi| \otimes |\chi\rangle\langle\chi|$, where $|\chi\rangle$ is a state of the remaining $(N-2)$ particles, and $|\tilde\Phi\rangle$ is a maximally entangled state equivalent up to $\sigma_z$ operations (which can be determined from the specific measurement outcomes) to $|\Phi\rangle$ (see Eq.~(\ref{Phi})). We emphasize that the operator ${\cal M}_j$ only acts non--trivially on particle $j$ and its neighbors. This is due to the fact (see {\bf (ii)}) that the operators $S_l^{j}$, $l=0,1,2,3$ --and thus the map ${\cal M}_j$-- only effect particle $j$ and/or its neighbors. It follows that in order to determine the reduced density operator of two neighboring particles $\{k,l\}$, $\rho_{kl}(t)$, one has to consider only the action of maps ${\cal M}_j$ which act on particles $k$,$l$ or neighbors of $k$ or $l$ on the maximally entangled state $|\Phi\rangle$, i.e.
\be
\label{rhoklgen}
\rho_{kl}(t)=(\prod_{j\in I}{\cal M}_j) |\Phi\rangle\langle \Phi|,
\ee
where $I=N_k \cup N_l \cup k \cup l$.  We have that the reduced density operator $\rho_{kl}(t)$ is distillable if and only if its partial transposition is non--positive \cite{Ho97}, i.e. $\rho_{kl}(t)^{T_k} \not\geq 0$.  To obtain a lower bound on the time until which the $N$--particle state $\rho(t)$ remains distillable one has to consider all neighboring pairs $\{k,l\}$, determine the corresponding threshold value on the lifetime of distillable entanglement $\kappa t_<^{kl}$ and take the minimum over all neighboring pairs $\{k,l\}\in E$ \cite{notedegree}. For graphs corresponding to periodic structures (e.g. some lattice geometry), such a minimization is however not required.

Thus we have that the threshold value is a function of the local degree (i.e. the number of neighbors) of the graph, but is independent of the number of particles $N$. Note, however, that the degree of the graph may itself depend on $N$ --as it is e.g. the case for GHZ states--, which then implies that the threshold value will indeed depend on $N$. In all cases where the degree of the graph is independent of $N$ --which is e.g. the case for all graphs corresponding to some lattice geometry, such as 2D/3D cluster state, hexagonal lattices, lattices with finite range interactions, etc.--, we have no scaling with $N$, i.e. the lower bound on the lifetime of entanglement is independent of the number of particles $N$. 
These results can also be understood in the following way: The measurement in the neighborhood of particles $k$ and $l$ disconnect these two particles from the remaining system, which implies that errors occurring in some outside area do not influence the two particles in question. This insight is also used in the following Sections and allows one to show that the behavior of cluster states is not a consequence of the specific decoherence model but rather a general feature of such states. 

The exact dependence of the distillability properties of $\rho_{kl}(t)$ (and thus the threshold value $p_<^{kl}$) on the graph $G$ can be determined as follows:
For $j \in N_k\setminus N_l$ the action of ${\cal M}_{j}$ can be described by a phase--flip channel acting solely on particle $k$, where a phase flip channel acting on particle $k$ is defined by
\be
\label{Mz}
{\cal M}_j^{(z)}\rho = p_z\rho +\frac{1-p_z}{2}(\rho+\sigma_z^{(k)}\rho\sigma_z^{(k)}),
\ee
and we find $p_z = 1- 2(p_1 + p_2)= 2(p_0+p_3)-1$. The action of ${\cal M}_{j}$ for $j \in N_l\setminus N_k$ can similarly be replaced by a phase--flip channel acting only on $l$. Moreover the action of ${\cal M}_j$ when particle $j \in N_k\cap N_l$ is a common neighbor of particles $k,l$ is given by a correlated phase--flip channel,
\be
{\cal M}_j^{(zz)}\rho = p_{zz}\rho +\frac{1-p_{zz}}{2}(\rho+\sigma_z^{(k)}\sigma_z^{(l)}\rho\sigma_z^{(k)}\sigma_z^{(l)}),
\ee
where $p_{zz}=1 - 2(p_1 + p_2)$. Note that the sequential application of each of these channels, say the correlated phase--flip channel with parameter $p_z$ for $|N_k\cap N_l|$ times, is equivalent to a single application of the same channel with new parameter $\tilde p=p^{|N_k\cap N_l|}$. Finally the Pauli channels ${\cal M}_{k}$ and ${\cal M}_{l}$ have also to be taken into account. In any case the resulting state $\rho_{kl}$ is diagonal in the ``Bell--basis'' $\{|\Phi\rangle,\eins\sigma_z|\Phi\rangle,\sigma_z\eins|\Phi\rangle,\sigma_z\sigma_z|\Phi\rangle\}$, where $\phi$ is given by Eq.~(\ref{Phi}). One can now easily determine $\rho_{kl}(t)$ for any graph $G$ and thus the condition when $\rho_{kl}(t)$ (Eq.~(\ref{rhoklgen})) has non--positive partial transposition and is thus distillable. After some algebra, one obtains that
the stated protocol yields distillable entanglement between $k$ and $l$ if 
\begin{itemize}
\item for the  depolarizing channel $ {\cal D}\rho =   p \rho + (1-p)\, \frac{1}{2}\eins $:
\be p^{|N_k|+1} +  p^{|N_k+N_l|} + p^{|N_l|+1} > 1 \label{lowerbDepol} \ee holds ,
\item for the  bit--flip channel $ {\cal D}\rho =   p \rho + \frac{1-p}{2} \left( \rho + \sigma_x \rho \sigma_x \right) $:
\be   p^{|N_k|} + p^{|N_k+N_l|} +  p^{|N_l|} > 1 \label{lowerbBitFlip}\ee holds,
\item for the phase--flip channel $ {\cal D}\rho =   p \rho + \frac{1-p}{2} \left( \rho + \sigma_z \rho \sigma_z \right) $:
\be \sqrt{2} - 1 < p \leq 1 \label{lowerbDephas}\ee holds,
\item for the quantum optical channel with $\mu=0$, i.e. $p_1=p_2=p$ and $p_3=q$ for $0\leq p,q < \frac{1}{4}$:
\begin{eqnarray} 
& & \left(1-2(p+q)\right) \big( (1-4p)^{|N_k|} +  (1-4p)^{|N_l|} + \cr 
& & \hspace{1cm} (1-4p)^{|N_k+N_l|-2}\left(1-2(p+q)\right) \big) > 1 
\end{eqnarray}  holds.
\end{itemize} 

A {\em lower bound on the lifetime} of distillable entanglement under decoherence due to one of the above Pauli channels can then be derived
 by solving the corresponding polynomial inequalities.
From Eq.(\ref{lowerbDephas}) it follows for example that in the case of the phase flip channel the lower bound obtained by this distillation protocol is the same for all graph states. This can be understood by the fact that here only the two individual dephasing channels acting on $k$ and $l$ (and not those of their neighbors) are relevant for the decoherence of the bell state $|\Phi\rangle$ between $k$ and $l$.
For the the bitflip and the depolarizing channel the critical value $p_<$ for $p$, which is proportional to the fidelity with the original pure graph state, increases with $|N_k|$, $|N_l|$ or $|N_k+ N_l|$. Similarly, the critical values for $(p,q)$ decrease with $|N_k|$, $|N_l|$ or $|N_k+ N_l|$, since $p$ and $q$ now represent the error probabilities instead of the fidelity with original pure graph state.  
In the following we will consider the condition (\ref{lowerbDepol})  for the depolarizing channel with $p=e^{-\kappa t}$ in more detail: 
\begin{figure}[th]
\includegraphics[width=8cm]{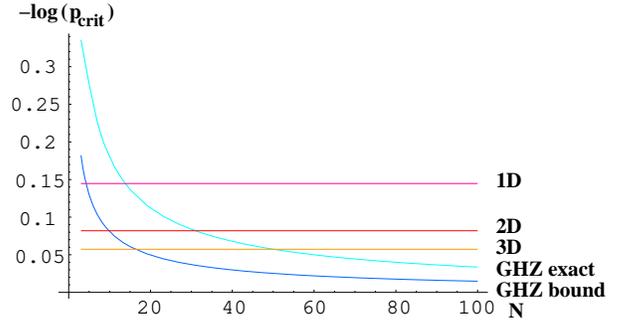}  
\caption{\label{fig:Lower} Under individual coupling due to the same depolarizing channel the lower bounds on $\kappa t$ to the lifetime of distillable $N$-party entanglement for the 1D- , 2D-, 3D-cluster state remain constant for arbitrary system sizes $N$ . For the $N$-party GHZ state the lower bound as well as the exact value for $\kappa t$ according to Eq.~(\ref{cond}), until which GHZ state remains distillable entangled, strictly decreases and goes to zero as $N\rightarrow \infty$.} 
\end{figure}
In the case of the $N$-party GHZ state both representations of Fig.~\ref{fig:GHZ} yield to the same polynomial inequality $2p^N + p^2>1$, since the depolarizing channel is invariant under local unitaries. This polynomial inequality can be further estimated from above giving $\kappa t > \frac{log(2)}{N}=\frac{0.6931}{N} $ As depicted in Fig.~\ref{fig:Lower} the corresponding critical value for $\kappa t$ is indeed always below the exact value given in
 Eq.~(\ref{cond}) and decreases with the number of particles $N$.
For linear chains or rings ($|N_k|=|N_{k+1}|=2$, $|N_k+N_{k+1}|=4$) one finds a threshold value $p_<=0.7167$, which gives a lower bound on the lifetime $\kappa t_< = 0.3331$. That is, for $p \geq p_<$ $(\kappa t \leq \kappa t_<)$ the state $\rho(t)$ is certainly $N$--party distillable using this specific protocol.
For cluster states corresponding to a regular 2D [3D] lattice we have that neighboring particles $k,l$ corresponding to inner vertices, i.e. with $|N_k|=|N_l|=4, |N_k+ N_l|=8$  $\left[ |N_k|=|N_l|=6, |N_k+ N_l|=12\right] $, will give the most sensitive polynomial inequality $p^5(p^3 + 2)>1$ $\left[ p^7(p^5 +2)>1 \right]$.  One hence finds $p_{<}=0.8281$ ($\kappa t_{<}=0.1886$) [$p_{<}=0.8765$ ($\kappa t_{<}=0.1318$)] respectively. As it can be seen in Fig.~\ref{fig:Lower}, whereas the lifetime of $N$-party GHZ states decreases
with the size of the system, cluster states do not show such a scaling behavior, since the derived lower bounds for cluster states remain constant.\\
 In the following we derive a more handy expression for the critical values for $p$ and $\kappa t$: For
 fixed degrees $|N_k|$ and $|N_l|$, one finds that the strongest lower bound on the lifetime (which is thus also valid for all other configurations of this kind) is obtained for $|N_k \cap N_l|=0$. This can be understood as follows. Assume that for some given graph one changes the graph such that the degree of two neighboring vertices $k,l$ increases by $1$, i.e. $|\tilde N_k| = |N_k|+1$ and $|\tilde N_l| = |N_l|+1$. The first possibility is that this increase is due to the addition of a single common neighbor of $k$ and $l$, i.e. $|\tilde N_k \cap\tilde N_l|= |N_k \cap N_l|+1$, which leads to the condition for distillability $p^{|N_k + N_l|}+p^{|N_k|+2}+p^{|N_l|+2} > 1$. In the second possibility the neighborhood of both particles $k$ and $l$ is increased by two different particles, i.e. $|\tilde N_k + \tilde N_l|=|N_k +N_l|+2$. In this case one obtains the condition
 $p^{|N_k+N_l|+2}+p^{|N_k|+2}+p^{|N_l|+2} > 1$ for distillability. Clearly, the second condition will give a larger value on $p$ and thus provides a stronger bound on the lifetime of distillable entanglement. Intuitively, this can be understood from the fact that adding a single joint neighbor corresponds to a {\em single} additional noise channel with correlated phase noise, while adding two independent neighbors corresponds to {\em two} independent noise channels acting on particles $k$ and $l$. The influence of two independent noise channels is larger than of a single (correlated) noise channel.
 In order to derive a lower bound we may therefore evaluate the polynomial inequalities of the different neighboring particles $k,l$, as if the value $|N_k+N_l|$ was maximal
(i.e. $|N_k\cap N_l|=0$), since this will give stronger/larger critical value than the critical value, which would be the solution to the exact polynomial inequality .
 Under this simplification and by 
using, that $p^{|N_k|+1}+p^{|N_l|+1} \geq 2p^{(|N_k|+|N_l|+2)/2} = 2p^{(|N_k + N_l|+2)/2}$, one finds that for 
\be
p > 2^{-\left(\frac{2}{|N_k|+|N_l|+2}\right)},
\ee
the reduced density operator $\rho_{kl}$ is certainly distillable. This leads to the lower bound on lifetime
\be
\kappa t_<= \frac{2\ln{(\frac{1}{2})}}{|N_k|+|N_l|+2}.
\ee
Taking $|N_k|$ and $|N_l|$ to be the maximal degrees of two neighboring vertices in the graph, this leads to a universal lower bound for all graph states under depolarizing noise.

We remark that the observed behavior, i.e. that the lifetime of multiparticle entanglement for cluster (and similar) states is essentially independent of
 the size of the system, also holds for more general decoherence models. This follows from the fact that ---similar to (ii)--- the action of {\em any} CPM acting on graph states describing an arbitrary decoherence process can be estimated by a CPM whose Kraus operators only contain products of $\sigma_z$ operators and the identity and thus the measurement (i) still commutes with the CPM. To this aim, we apply after the application of the CPM a local depolarization procedure which maps arbitrary density operators to operators diagonal in the graph state basis without changing the diagonal elements \cite{Du04}. When restricted on graph states, the resulting action of the initial CPM (given by $\sum_{k,l} a_{k,l} O_k \rho O_l$, where $O_k,O_l$ are products of Pauli operators)) can then be described by a CPM specified by $\sum_k a_{k,k} O_k \rho O_k^\dagger$, where all operators $O_k$ can be expressed in terms of products of $\sigma_z$ operators. 
 Only operators $O_k$ which act non--trivially on particles $k,l$ or their neighbors in the graph affect the resulting maximally entangled pair after the measurement (i), leading again to a threshold value which is independent of the size of the system for all those decoherence models where the number of such operators $O_k$ is independent of $N$. This is for instance the case if each $O_k$ acts non--trivially on a finite, localized number of subsystems.
Therefore, the fact that all graph states with finite maximal degree, such as cluster states, the lifetime of distillable $N$-party entanglement will remain finite, holds in particular for all decoherence models based on an {\it arbitrary individual coupling to the environment} described by a quantum optical channel in Eq.~(\ref{QO}). In the following we will determine {\it upper bounds} on the distillable entanglement.

\subsection{Upper bound I: Noise operation becomes entanglement breaking}\label{upper1}

In our first approach, we determine an upper bound to the lifetime of $N$-party entanglement by considering the capability of the decoherence process
 to disentangle any state disregarding its specific form.
 Hence the upper bounds derived in this way will apply not only to graph states but to an arbitrary state. In turn this method will be restricted to
coupling of the particles to individual environments described by an arbitrary channel of the form
 \be\label{indivNoise} \mathcal{D}\rho= \sum_{i,j =0}^3 \,p_{ij} \sigma_i \rho \sigma_j \ee like the
completely positive map (CPM) in Eq.(\ref{QO}).
We now make use of the Jamiolkowski isomorphism between CP maps and states \cite{Isomorphism}:
 Let $|\Phi^+\rangle^{kk'} = \frac{1}{\sqrt{2}} \left( |0\rangle^k |0\rangle^{k'} +| 1\rangle^k |1\rangle^{k'} \right)$
 denote the maximally entangled state on system $k$ and a copy  $k'$ of the system $k$. Then to each CPM $\mathcal{D}$ acting on particle $k$ there uniquely corresponds a
 state 
\be\label{Iso}
 \sigma_\mathcal{D}^{kk'}\equiv \mathcal{D}^k \left(|\Phi^+\rangle^{kk'}\langle\Phi^+ |\right)
\ee
 on the combined system of $k$ and $k'$.
The main fact which we will use in the following is, that \\
{\bf (i)} the CPM $\mathcal{D}$ is {\it entanglement breaking} \cite{Shor}, i.e. $\mathcal{D}^k\rho^{k A}$ is separable for any
(possibly entangled) state $\rho^{kA}$ on the composite system, consisting of particle $k$ and some other particles hold by the parties $A$, if and only if the corresponding state $\sigma_\mathcal{D}^{kk'}$ is separable (with respect to the particles $k$ and $k'$).\\
 Hereby the 'only if' implication directly follows from the very definition of 
the map to be completely disentangling, whereas the other direction can be seen as follows: Given the state $\sigma_\mathcal{D}^{kk'}$ one obtains the
 corresponding CPM via the inverse isomorphism, i.e. 
\be\label{IsoInverse}
\mathcal{D} \rho^{k_0} \equiv\; 2^4\times\,\text{tr}_{kk_0}\left(|\Phi^+ \rangle^{kk_0}\langle \Phi^+|\, \rho^{k_0}\otimes \sigma_\mathcal{D}^{kk'} \right) \; ,
\ee
where $\rho$ is an arbitrary state on another copy $k_0$ of system $k$,
the projection onto $|\Phi^+\rangle^{k_0k}$ is performed with respect to the joint system $k_0,k$ and $\mathcal{D}$ is
 now thought to map system $k_0$ onto system $k'$ instead of $k$ onto $k$.
Now, if $\sigma_\mathcal{D}^{kk'}= \rho_1^k \otimes \rho_2^{k'}$ is separable, then (\ref{IsoInverse}) factorizes into 
\be 
 ^{kk_0}\hspace{-0.05cm}\langle\Phi^+| \rho^{k_0}\otimes \rho_1^{k} |\Phi^+ \rangle^{kk_0} \otimes \rho_2^{k'}
\ee 
disregarding, whether$\rho^{k_0}=\rho^{k_0 A}$ is itself entangled with some other parties $A$ or not. The resulting state on system $k'$ which corresponds to
 $\mathcal{D}\rho$ is even independent of the input state $\rho$ and thus cannot be entangled with the parties $A$ whatsoever. 

In order to derive an upper bound to the entanglement of states suffering from decoherence due to individual coupling $\mathcal{D}^k$ of the particles to the environment,
 one can determine the critical value for $p_{ij}$ in (\ref{indivNoise}), for which the state
\be \label{RhoD}
 \sigma_{\mathcal{D}^k} = \sum_{i,j =0}^3 \,p_{ij}\, \sigma^k_i |\Phi^+\rangle\langle \Phi^+| \sigma^k_j =  \sum_{i,j =0}^3 \,p_{ij}\, |\Phi_i\rangle\langle \Phi_j|
\ee
becomes separable and hence the CPM $\mathcal{D}^k$ become entanglement breaking. In (\ref{RhoD}) we have used the notation 
$|\Phi_i\rangle = \sigma_i | \Phi^+ \rangle$, where $(|\Phi_0\rangle,|\Phi_1\rangle,|\Phi_2\rangle,|\Phi_3\rangle)$ form a complete 'Bell'-Basis.
In the following we will restrict to the same individual coupling $\mathcal{D}^k=\mathcal{D}$ of the particles to the environment, which then only requires to test the
 separability of one state $\sigma_{\mathcal{D}}$. In the case of Pauli channels $\mathcal{D}\rho = \sum_{i=0}^3 \,p_{i}\, \sigma_i \rho \sigma_i$ this task becomes particularly easy, since the state $\sigma_{\mathcal{D}}$ is diagonal in the above 'Bell' basis. Moreover for such Bell diagonal states the separability criterion reduces to the
 necessary and sufficient condition, that all diagonal entries $p_i$ are smaller than $1/2$, i.e.
\be \label{upper1Pauli}
\max_{i=0,1,2,3}\, p_i \leq \frac{1}{2} \, .
\ee
This can be easily evaluated for the examples given in Sec.~\ref{individual}.
For the depolarizing channel $ {\cal D}\rho =   p \rho + (1-p)\, \frac{1}{2}\eins $ the state $\rho(t)$ has certainly become $N$-party separable, if
 \be p\leq p_> \equiv \frac{1}{3}\; .\ee
Note that this condition provides a universal upper bound for {\it all} states exposed to individual depolarizing channels.
For the quantum optical channel with $\mu=0$ in Eq.~(\ref{QO}) one arrives at the condition
$
2\lambda_1 + \lambda_3 \geq\frac{1}{2}
$.\\
In the case of a general quantum optical channel with $\mu\not= 0$ or an arbitrary noise channel of the form (\ref{indivNoise}) one can instead use the fact, that for any
 two dimensional systems $k$ and $k'$ the PPT [NPT] criterion, i.e. the positivity [non-positivity] of the partial transpose $\rho^{T_k}$, is necessary and sufficient for 
separability [distillability] \cite{Pe96,Ho96}. Thus the CPM $\mathcal{D}$ is entanglement breaking, if and only if
\be
\left(\rho^{kk'}_{\mathcal{D}}\right)^{T_k} \, \geq \, 0\; .
\ee
For the general quantum optical channel (\ref{QO}) this leads after some algebra to the condition
\be
\lambda_1^2 - \mu^2\geq \left(\lambda_1 + \lambda_3 -\frac{1}{2}\right)^2\, .
\ee
In terms of the original parameters $B, C$ and $s$ of the quantum optical master equation with the superoperator defined in Eq.~(\ref{Lindblad}) this inequality reads
\be \label{QO_ent_breaking}
s(1-s)\,\left[e^{C t}\left( 1 - e^{-B t} \right) \right]^2  \geq  1 \; .
\ee
It is worth remarking, that in the terminology of quantum optics (reservoir theory) both the equilibrium value $s$ and the decay rates $B$, $C$ enter in the inequality in this multiplicative form.
For the example of a decay channel i.e. $\lambda\equiv \lambda_1 =\lambda_2 = \mu$, we have
$
 0 \geq \left(\lambda + \lambda_3 -\frac{1}{2}\right)^2\, ,
$
which cannot be satisfied. Therefore the decay channel cannot become entanglement breaking and the multi party GHZ states are an example for states, that remain entangled under decoherence due to this channel (see Sec.~\ref{GHZ_QO}).
For the  bitflip channel $ {\cal D}\rho =   p \rho + \frac{1-p}{2} \left( \rho + \sigma_x \rho \sigma_x \right) $ and the dephasing channel $ {\cal D}\rho =   p \rho + \frac{1-p}{2} \left( \rho + \sigma_z \rho \sigma_z \right) $ the upper bound $p=0$ obtained from Eq.~(\ref{upper1Pauli}) becomes trivial.

\subsection{Upper bound II: Noisy Ising interaction becomes separable}\label{upper2}

In our second approach, we determine an upper bound $t_>$ on the lifetime of distillable entanglement by showing that after a certain time, the state $\rho(t)$ becomes fully separable and is hence no longer entangled whatsoever. To this aim, we consider the (dynamical) description of graph states in terms of Ising interactions acting on a specific separable state. We determine the separability properties of the operator $\rho(t)$ by considering the corresponding interactions which generate the state and show that for a given noise level, these operations itself become separable and hence are not capable of creating entanglement. Consequently, also the state $\rho(t)$ is separable in this case. The main advantage of this approach is that one does not have to consider the $N$--particle state $\rho$ itself and determine when it is fully separable (a task which is generally very difficult, especially if $N$ is large), but has to consider only {\em two--particle operations} and determine when these operations are separable.

We make use of the following properties: 

{\bf (i)} The graph state $|G\rangle$ corresponding to a graph $G$ can be written as (see Sec.~\ref{definitions}) \cite{Rau01}
\be
|G\rangle =\prod_{\{k,l\}\in E}U_{kl}|+\rangle^{\otimes N},
\ee
where $U_{kl}\equiv e^{-i\pi (\mathbf{1}^{(k)}-\sigma_z^{(k)})/2\otimes(\mathbf{1}^{(l)}-\sigma_z^{(l)})/2}$ and $|+\rangle=1/\sqrt{2}(|0\rangle+|1\rangle)$.

{\bf (ii)} We will only take $\sigma_z$- noise into account. We thus restrict the following analysis to decoherence models due to the same individual noise channel $ \mathcal{E}$, that can be  decomposed into some noise channel $\mathcal{E'}$ acting after a dephasing channel $\mathcal{D}\rho =p_z\rho+\frac{1-p_z}{2}[\rho+\sigma_z\rho\sigma_z]$,
 i.e. $\mathcal{E}=\mathcal{E'}\circ \mathcal{D}$. A more detailed analysis of the cases, for which such a decomposition is possible, is postponed to the appendix A.
For the depolarizing channel ${\cal D}_k(p)$ (Eq.~(\ref{Depol}) with noise parameter $p\equiv e^{-\kappa t}$), such a decomposition is possible choosing $p_z=\frac{2p}{1+p}$
and
\be\label{Depolp_z}
\mathcal{E'}_k \rho = \frac{1+p}{2}\rho+\frac{1-p}{4} \left[\sigma_x^{(k)}\rho\sigma_x^{(k)} + \sigma_y^{(k)}\rho\sigma_y^{(k)}\right] \; .
\ee
 This can be checked by direct calculation. 

We now investigate the influence of noise on the entanglement generating unitary operation $U_{kl}$ and determine when the resulting CPM becomes separable. Since $U_{kl}$ commutes with ${\cal D}_j(p_z)$, it follows that $\rho(t)$ can be written as
$
\rho(t)= \mathcal{E'}( \tilde\rho(t))
$,
where $\tilde \rho(t)$ is obtained from the original graph state by considering only phase noise described by ${\cal D}_k (p_z)$, i.e
\be
\tilde\rho(t) \equiv \prod_j {\cal D}_{j}(p_z) 
\prod_{\{k,l\}\in E}  U_{kl}|+\rangle\langle+|^{\otimes N} U_{kl}^\dagger,
\ee
Since $\rho(t)$ is obtained from $\tilde\rho(t)$ by means of separable operations, it is sufficient to determine the condition when $\tilde\rho(t)$ becomes separable. In principle, one could also consider this additional noise to obtain a stronger upper bound on the lifetime, however the analysis becomes more involved in this case as one has to deal with correlated noise. In the following we will therefore consider only noise resulting from phase--flip errors described by ${\cal D}_k$, i.e. the map
\be
\label{Dz}
\tilde {\cal D}_{kl}(p_z,q_z) \rho \equiv {\cal D}_k(p_z) {\cal D}_l(q_z) U_{kl}\rho U_{kl}^\dagger\; ,
\ee
 for two different dephasing parameters $p_z$ and $q_z$. With this notation the operator $\tilde\rho(t)$ can be written as 
\be
\label{tilderho}
\tilde\rho(t)= \prod_{(k,l)\in E} \tilde {\cal D}_{kl}(p_z^{1/{|N_k|}},p_z^{1/{|N_l|}}) |+\rangle\langle+|^{\otimes N}.
\ee
For the vertex $k$ with degree $|N_k|$ we have split up the action of the map ${\cal D}_{k}(p_z)$ into $|N_k|$ parts (one for each term in the product which involves
 $U_{kl}$ and thus particle $k$) by using a decomposition of the the map ${\cal D}_k (p_z)$ into
\be
{\cal D}_k (p_z) \rho = \prod_{j=1}^m {\cal D}_k(p_z^{1/m}) \rho. \label{splitup}
\ee
 This leads to the parameter $p_z^{1/{|N_k|}}$ in (\ref{tilderho}).
If in Eq.~(\ref{tilderho}) all maps ${\cal D}_{kl}(p_z^{1/{|N_k|}},p_z^{1/{|N_l|}})$ at a fixed vertex $k$ are separable, it immediately follows that also $\tilde \rho(t)$ is
 $k$-versus-rest separable since the following maps are local and act on a $k$-versus-rest separable state.
 
To determine the entanglement properties of $\tilde {\cal D}_{kl}(p_z,q_z)$, we make again use of the Jamiolkowski isomorphism \cite{Isomorphism} between CPM and mixed states \cite{Ci00}. In particular, we use that a CPM ${\cal D}$ is separable and hence not able to generate entanglement if the corresponding mixed state $D$ is separable, where 
\be
D_{k_1k_2l_1l_2}={\cal D}_{k_2l_2} |\Phi^+\rangle_{k_1k_2}\langle \Phi^+| \otimes |\Phi^+\rangle_{l_1l_2}\langle \Phi^+|,
\ee
and separability has to be determined between parties $(k_1k_2)$ and $(l_1l_2)$. It turns out to be useful to define
\bea
|\phi_{00}\rangle &\equiv& \frac{1}{\sqrt{2}}(|\tilde0\rangle_{k_1k_2} |\tilde0\rangle_{l_1l_2} + |\tilde 1\rangle_{k_1k_2} |\tilde 1\rangle_{l_1l_2}), \nonumber\\
|\phi_{01}\rangle &\equiv& \frac{1}{\sqrt{2}}(|\tilde0\rangle_{k_1k_2} |\tilde1\rangle_{l_1l_2} + |\tilde 1\rangle_{k_1k_2} |\tilde 0\rangle_{l_1l_2}), \nonumber\\
|\phi_{10}\rangle &\equiv& \frac{1}{\sqrt{2}}(|\tilde0\rangle_{k_1k_2} |\tilde0\rangle_{l_1l_2} - |\tilde 1\rangle_{k_1k_2} |\tilde 1\rangle_{l_1l_2}), \nonumber\\
|\phi_{11}\rangle &\equiv& \frac{1}{\sqrt{2}}(|\tilde0\rangle_{k_1k_2} |\tilde1\rangle_{l_1l_2} - |\tilde 1\rangle_{k_1k_2} |\tilde 0\rangle_{l_1l_2}),
\eea
with
\bea
|\tilde 0\rangle_{k_1k_2} &\equiv& |00\rangle_{k_1k_2}, \nonumber\\
|\tilde 1\rangle_{k_1k_2} &\equiv& |11\rangle_{k_1k_2}, \nonumber\\
|\tilde 0\rangle_{l_1l_2} &\equiv& \frac{1}{\sqrt{2}}(|00\rangle_{l_1l_2} + |11\rangle_{l_1l_2}), \nonumber\\
|\tilde 1\rangle_{l_1l_2} &\equiv& \frac{1}{\sqrt{2}}(|00\rangle_{l_1l_2} - |11\rangle_{l_1l_2}).
\eea
One finds that the state $\tilde D_{k_1k_2l_1l_2}(p_z,q_z)$ corresponding to the map $\tilde {\cal D}_{kl}(p_z,q_z)$
 is given by
\be
\tilde D_{k_1k_2l_1l_2}(p_z,q_z) = \sum_{i,j=0}^1 \lambda_{ij} |\Phi_{ij}\rangle\langle\Phi_{ij}|,
\label{Dzstate}
\ee
where $\lambda_{00}=(1+p_z)(1+q_z)/4, \lambda_{01}=(1+p_z)(1-q_z)/4,\lambda_{10}=(1+q_z)(1-p_z)/4, \lambda_{11}=(1-p_z)(1-q_z)/4$.
This state is separable with respect to $(k_1k_2)-(l_1l_2)$ if and only if $\lambda_{00} \leq 1/2$, as its partial transposition is positive in this case.
Note that for systems in $\C^2\otimes \C^2$, positivity of the partial transposition is a sufficient condition for separability \cite{Pe96,Ho96}. 
Although the system that we consider consists of two four--level systems, the resulting state has support only in a four dimensional subspace and
 thus the results about qubit systems can be directly applied. We then obtain that the operator $\tilde D_{k_1k_2l_1l_2}(p_z,q_z)$ is separable --and hence
 the CPM $\tilde {\cal D}_{kl}(p_z,q_z)$ is separable and not capable to create entanglement-- if and only if
\be
\label{condsep}
(1+p_z)(1+q_z) \leq 2.
\ee

We now use the above result to obtain an upper bound for the lifetime of graph states under a decoherence model, that obeys (ii).
Due to Eq.~(\ref{condsep}) we find that the map ${\cal D}_{kl}(p_z^{1/{|N_k|}},p_z^{1/{|N_l|}})$ is separable if
 \be (1+p_z^{1/|N_k|})(1+p_z^{1/|N_l|}) \leq 2\; .\ee 
The threshold value $p_>$ such that state $\rho(t)$ is fully separable is then obtained by considering all pairs of particles $\{k,l\}$, calculate the corresponding value 
$p_>^{kl}$ and take the minimum over all $\{k,l\}$, i.e. $p_>= \min p_>^{kl}$. This ensures that all involved operators are separable for $p_z \leq p_>$. 
By estimating $|N_k|$ and $|N_l|$ from above with maximal degree $m$ in the graph, we arrive at the weaker upper bound \be p_z \leq (\sqrt{2}-1)^m\; .\ee
For the various decoherence processes, one now has to determine the actual value for $p_z$, which depends on the parameters of the underlying noise model and should be chosen minimal (see appendix A), since this gives the strongest upper bound.
 The exact values for the bounds obtained in this way are however worse than the upper bound derived in Sec.
~\ref{upper1}, but as we will see in Sec.~\ref{weighted} the way of deriving the upper bound here will turn out to be well suited for all those cases where the initial state is only slightly entangled.    

Moreover, as it was the case for the lower bound, the derived upper bound on the lifetime of distillable entanglement does only depend on the maximum degree of the graph and not necessarily on the number of particles $N$. 
We remark that the upper and lower bound on the lifetime of graphs states show a different dependence on the degree $m$ of the graph. While the lower bound on the lifetime decreases with $m$, the upper bound on the lifetime increases with $m$. We emphasize that this observation applies only to the lower and upper bounds, and no definitive statement about the actual dependence of the lifetime of distillable entanglement on the degree of the graph can be made (although one may expect that the lifetime of entanglement decreases with the degree of the graph). The different dependences of the lower and upper bound can in part be understood by looking at the corresponding derivations. In particular, in the derivation of the upper bound $t_>$ the influence of both $\sigma_x$ and $\sigma_y$ noise is completely ignored. The influence of this kind of noise, however, strongly depends on the degree of the graph and is in fact responsible that e.g. the fragility of GHZ states depends on the number of particles \cite{Du04}. That is, $\sigma_x$ noise on all neighboring particles acts as $\sigma_z$ noise on a given vertex, and the noise accumulates. However, it is not straightforward to take also ${\cal E}_x^{(k)}$ and ${\cal E}_y^{(k)}$ in above analysis into account, as they lead to correlated noise when expressed in terms of $\sigma_z$ operators (see Eq.~(\ref{Mk})). This implies that one could no longer consider separability properties of two--qubit maps independently but has to take correlations into account and thus consider a larger (or eventually the whole) system, thereby losing the main advantage of this approach on determining separability of the resulting state.

\subsection{Upper bound III: Partial transposition criterion for graph diagonal states}\label{upper3}

In our third approach, we determine an upper bound on the lifetime by considering the partial transposition with respect to several partitions. Although this upper bound will be worse than the upper bound in Sec.~\ref{upper1} (except for some singular cases), the ability to explicitly compute the partial transpose with respect to different partitions will enable us to compare the afore mentioned bounds with the exact critical values for the PPT criterion, at least for graphs with only few vertices 
$(N\leq 10)$. In this case the techniques developed in this Sec. thus lead to stronger results for the lifetime of $N$--party entanglement. For the following upper bound we make use of the fact that a $N$--particle state is certainly no longer $N$--party distillable if at least one of the partial transpositions with respect to all possible bipartite partitions is positive. 
To this aim, we determine the eigenvalues of partial transposition of $\rho(t)$ with respect to various partitions.
Since this is in general a rather complicated task, we will assume that decoherence of the particles is based on the same individual Pauli channel:
\begin{equation}
 {\cal D}_{k}\rho =  \sum_{j=0}^3 p_j(t)\sigma^{k}_j \rho \sigma^{k}_j \; .
\end{equation}
As it was already used in the derivation of the lower bound, under such a Pauli channel the graph state $|G\rangle$ evolves in time into a mixed state 
$\rho \equiv \prod_{k \in V} {\cal D}^{(k)}\,|G \rangle \langle G |\; ,$
that is diagonal in the graph state basis $|U\rangle_G $:
\begin{equation}\label{RhoG}
\rho =  \; \sum_{U\subseteq V} \, \lambda_{U}\,  |U\rangle_G \langle U  | \, . 
\end{equation} 
In the following we will make use of the following facts, whose proofs are postponed to the appendix B:\\

{\bf (i) } The diagonal elements $\lambda_U$ in Eq.~(\ref{RhoG}) can be computed to be of the form 
\begin{eqnarray}
\label{PauliLambda}
\lambda_U & = & p_0^{|V|} \,  \sum_{U'\subseteq V} \, q_1^{|U'\setminus (\Gamma U' + U)|}\, q_2^{|U'\cap (\Gamma U' + U)|} \, \times \cr 
& &  \hspace{1,5cm}\times \, q_3^{| (\Gamma U' +U)\setminus U'|}\; , 
\end{eqnarray}
where $q_i:=\frac{p_i}{p_0}$ for $i=1,2,3$.
In the case of the depolarizing channel $(q:=q_1=q_2=q_3=\frac{1-p}{3p+1})$ this simplifies to
\begin{equation}
\label{DepolLambda}
\lambda_U = p_0^{|V|} \,  \sum_{U'\subseteq V} \, q^{|U'\,\cup\, (\Gamma U' +U)|}\; .
\end{equation}
We note, that in both expressions we have made use of the notational simplifications described in Sec.~\ref{definitions}. \\

{\bf (ii) } For any state $\rho$ of the form (\ref{RhoG}), i.e. that is diagonal in the basis $|U\rangle_G$ according to some graph $G$, the 
partial transposition $\rho^{T_A}$ with respect to some partition $A$ is again diagonal in the (same) graph state basis $|U\rangle_G$.
In order to compute the corresponding eigenvalues, let $\Gamma'=\Gamma_{AA^c}$ denote the adjacency matrix of the graph between the partition $A$
and its complement $A^c$, i.e.
\begin{equation}\label{Gamma for bi-partition}
\left(\begin{array}{cc}
  \Gamma_{A} & \Gamma_{AA^c}^T \\
   \Gamma_{AA^c} & \Gamma_{A^c} \\
\end{array}\right)
= \Gamma .
\end{equation} 
Then:
\begin{eqnarray} 
\label{PTofRhoG}
  \rho^{T_A}  & = &  \sum_{U \subseteq V} \, \lambda'_U \, |U \rangle_G\langle U|  \; \; \text{with}\\  
\lambda'_U & = &  \frac{|\text{ker}\,\Gamma'|}{2^{|A|}}\, \sum_{(X,Y) \in \atop (\text{ker}\,\Gamma')^{\bot} \times (\text{Im}\,\Gamma')} \, 
(-1)^{\langle X , A_Y  \rangle} \, \lambda_{\left( U + X + Y\right)} \nonumber \; ,
\end{eqnarray}
where $A_Y \in A$ is arbitrary with $\Gamma' A_Y = Y$ 
 and the kernel $\text{ker}$ or the orthocomplement $\bot$ are taken with respect to the subspace $\mathcal{P}(A)$ spanned by the sets in $A$.\\

{\bf (iii) } In the case of small noise $0< q_i =\frac{p_i}{p_0} \leq 1 $  for $i=1,2,3$ the 
following estimation:
\begin{equation}\label{LambdaEstimation}
q\,\lambda_U \leq \lambda_{U+k} \leq \frac{1}{q}\, \lambda_U\; ,
\end{equation}
can be derived, where $q=\min (q_1,q_2,q_3)$. The same holds for $\lambda_{U+N_k}$ and $\lambda_{U+N_k+k}$ instead of $\lambda_{U+k}$.\\

Before coming to the upper bound let us give two examples for formula (\ref{PTofRhoG}):
If $\Gamma'$ is invertible, then $\text{ker}\, \Gamma' =\{0\}$ and $(\text{ker}\, \Gamma')^\bot =\mathcal{P}(A)$ holds.
 Moreover $(\ref{PTofRhoG})$ can be simplified by parameterizing $\text{Im}\,\Gamma'$ with $Y=\Gamma' A_2$, where $A_2 \subseteq A$:
\be
\lambda'_U  =   \frac{1}{2^{|A|}}\, \sum_{A_1,A_2 \subseteq A} \, 
(-1)^{\langle A_1 , A_2  \rangle} \, \lambda_{\left( U + A_1 + \Gamma' A_2\right)} \; .\ee
If $A=\{k\}$ for a non-isolated vertex $k\in V$ the eigenvalues of the partial transposition with respect to $A$ are
\begin{equation}\label{PTofa} 
\lambda'_U  =   \frac{1}{2}\, \left( \lambda_U + \lambda_{U + N_k} + \lambda_{U + k} - \lambda_{U + N_k + k}\right) \; .
\end{equation}
Similarly for the partial transposition with respect to the split $A=\{k,l\}$ versus rest, where $k,l \in V$  are two non-adjacent vertices
 with linearly independent neighbor sets $N_k$ and $N_l$, one obtains:
\begin{equation}\label{PTofab}
\lambda'_U   =   \frac{1}{4}\, \left( \sum_{X \in \mathcal{M}_+} \, \lambda_{U+X} - \sum_{X \in \mathcal{M}_-} \, \lambda_{U+X} \right) \; , 
\end{equation}
where
\begin{eqnarray} 
 \mathcal{M}_+ & = & \{ 0, k, l, k+l, N_k, N_l, N_k+N_l, k+N_l, \cr & & l+N_k, k+l+N_k+N_l\}\; \;\text{and} \cr
\mathcal{M}_- &= &\{ k+ N_k,l+ N_l, k+N_k+N_l, l+N_k+N_l, \cr & &   k+l+N_k, k+l+N_l\} \; .\nonumber
\end{eqnarray}
If $k$ and $l$ are adjacent the same formula holds but with neighbor sets $N'_k=N_k\setminus l$ and $N'_l=N_l\setminus k$ restricted to $A^c$.\\
Finally we note, that for GHZ diagonal states of the form (\ref{rhot}) the positivity [non-positivity] of the partial transpose with respect to all possible partitions
 was already a necessary and sufficient condition for $N$-party separability [distillability]. For general graph diagonal states the corresponding PPT [NPT] criterion is
 only known to be a necessary condition for $N$-party separability [distillability], whereas the sufficiency of these conditions is presently unknown. 
But for all partitions $(A,A^c)$, for which the pure graph state $|G\rangle$ has Schmidt measure $1$, i.e. it can be decomposed into the form
 $|G\rangle = \alpha_1 |a_1\rangle^A |b_1\rangle^{A^c} + \alpha_2 |a_2\rangle^A |b_2\rangle^{A^c}$, the NPT criterion is also sufficient condition 
at least for the distillability of a $(A,A^c)$-entangled state: If $\rho^{T_A}$ in Eq.~(\ref{PTofRhoG}) has a negative eigenvalue $\lambda'_U$, then the 
 corresponding eigenstate $|U\rangle_G$ has a Schmidt decomposition of the form  $|U\rangle_G= \sigma_z^U |G\rangle = 
\alpha_1 |a'_1\rangle^A |b'_1\rangle^{A^c} + \alpha_2 |a'_2\rangle^A |b'_2\rangle^{A^c}$ and also a negative overlap $_G\hspace{-0.05cm}\langle U| \rho^{T_A}| U\rangle_G < 0$,
 which is sufficient for  $(A,A^c)$-distillability \cite{Ho98}.

These results can now be used to derive upper bounds to the distillable entanglement in graph states in the presence of local noise described by a Pauli channel $(\ref{PauliChannel})$ with $p_i>0$ for $i=1,2,3$.
For example, if one considers the split one-versus-rest, the eigenvalues $\lambda'_U$ of the partial transposition with respect to the corresponding partition $A=\{k\}$ in 
Eq.~(\ref{PTofa}) can be bounded from below by 
\be \label{estim} \lambda'_U \geq (1 + 2q - \frac{1}{q}) \, \lambda_U \; .\ee
Therefore the state $\rho$ in $(\ref{RhoG})$ is certainly PPT with respect to the partition $A=\{k\}$ if $1 + 2q - \frac{1}{q}\geq 0$, i.e. $\frac{1}{2}\leq q\leq 1 $.
 In the case of the depolarizing channel $(q=\frac{1-p}{3p+1})$ this means, that no $A-A^c-$ entangled state can be distilled from any graph state $\rho$ if
$p(t)$ falls below $p_>=\frac{1}{5}$.\\
For the partial transposition with respect to the partition $A=\{k,l\}$ (see $(\ref{PTofab})$), Eq. $(\ref{LambdaEstimation})$ can be applied twice (e.g. 
 $\lambda_{U+k+N_l}\geq q \lambda_{U+k} \geq q^2 \lambda_{W}$ ) in order to obtain estimations for the 'higher order' terms $k+l, k+N_l , k+l+N_k+N_l \ldots $
in $\mathcal{M}_+$ and $\mathcal{M}_-$. In this way one arrives at the condition
\be \lambda'_U \geq (1+4q+5q^2-\frac{2}{q}-\frac{4}{q^2})\, \lambda_U \geq 0 \ee
 for the distillability of $A-A^c-$entanglement. This means, that in the case of the depolarizing channel for $q\geq 0.8457$ or $p\leq 0.0436$  any graph state $|G\rangle$ will become PPT with respect to $A=\{k,l\}$.
A closer comparison with Sec.~\ref{upper1} shows, that the upper bounds to $N$-party distillable entanglement
 derived in this way are worse than the upper bound in Eq.~(\ref{upper1Pauli}). 
For the one-versus-rest split, this can be understood for a general Pauli channel with $p_i>0$  by rewriting the condition as 
\be \label{upper2Pauli}
\min_{i=1,2,3}\, p_i \geq \frac{p_0}{2} \; .
\ee 
Then, due to $q\geq 1$, $p_0$ must be the maximum in Eq.
 (\ref{upper1Pauli}) and hence can only be larger than $\frac{1}{2}$ if also $\frac{p_0}{2}\geq \frac{1}{4}$ holds, which cannot be exceeded by the minimum in 
(\ref{upper2Pauli}).  Nevertheless, we think that the derivation can be of interest for other applications involving the partial transposition of graph diagonal states. 
In particular, it is an open question, whether for certain Pauli channels with $p_i>0$ the conditions for PPT with respect to larger partitions might yield a stronger upper bound than the condition in Eq.~(\ref{upper1Pauli}). This will certainly depend on the solutions to the corresponding polynomial equation in $q$.
 But any upper bound derived with the use of (iii) will not depend on the topology of the underlying graph in question. 
By using a slight modification of the argumentation leading to the estimation in (iii) we will therefore discuss the example of the dephasing channel, 
for which a stronger upper bound can be provided, that conversely depends on the topology of the graph.
In any case, the procedure to compute the eigenvalues of the partial transposition described in (iii) does not require the diagonalization of a $2^N\times2^N$-matrix and therefore allows the evaluation of the PPT criteria with respect to different partitions, as long as the vector consisting of the initial eigenvalues $\lambda_U$ (which is already of length $2^N$) is small enough to be stored and -in the case, that it occurs as a result of Pauli channel- as long as this vector can also be initialized fast enough. In order to illustrate the afore mentioned results we have, for example, considered rings up to size $N=10$ suffering from decoherence due to the depolarizing channel and examined the partial transpose with respect to all possible partitions.
 Fig.~\ref{fig:DepolRing} depicts the critical value for $p$, after which the state $\rho$ first becomes PPT with
 respect to some partition, which implies that at this point the state $\rho$ is certainly no longer $N$-party distillable. For  Fig.~\ref{fig:DepolRing} the critical value 
$p_{\text{crit}}$ has also been computed, after which the state $\rho$ has become PPT with respect to all partitions, i.e. after which $\rho$ contains at most
 bound entanglement with respect to any partition.  In contrast to the case of $N$-party GHZ states, 
for which the one-versus-rest partition is the first to become PPT, the numerical results for small $N$ indicate that in rings this split seems to be most 
stable against decoherence due to noise described individual depolarizing channels and that the smallest eigenvalue of the partial transposition with respect to these one-versus-rest splits $\{k\}$ is given by $\lambda_{N_k+k}$.
\begin{figure}[th]
\includegraphics[width=8cm]{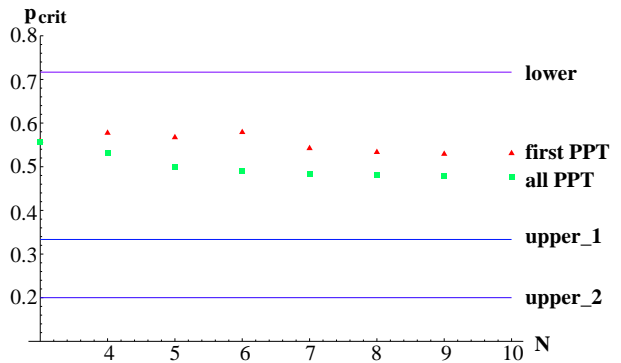}  
\caption{\label{fig:DepolRing} For the case of particles in rings of size $N\leq 10$, which individually decohere according to the same depolarizing channel (\ref{Depol})
 with parameter $p$:  the critical value $p_{\text{crit}}$, after which the first [last] partition becomes PPT $\triangle$ [$\Box$], the lower bound according 
to Sec.~\ref{lower} and the upper bounds according to Sec.~\ref{upper1} and Sec.~\ref{upper3}. }
\end{figure}

In Fig.~\ref{fig:DepolLU} we show representatives of the equivalence classes for connected graphs over $N=5,6,7$ vertices discussed in \cite{He03},
 that are most stable or instable, when exposed to noise described by individual depolarizing channels.
\begin{figure}[th]
\includegraphics[width=9cm]{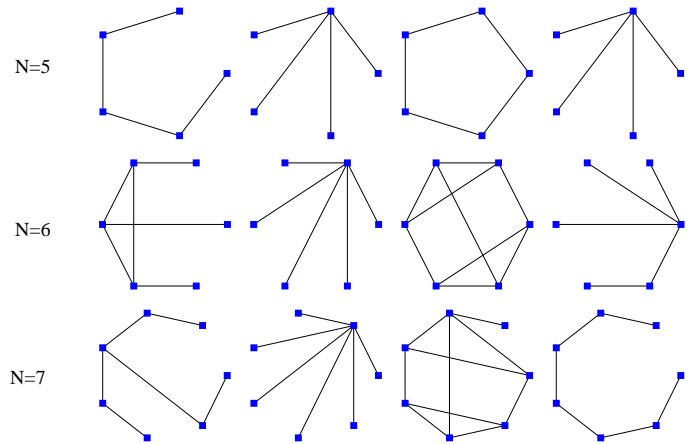}  
\caption{\label{fig:DepolLU} Representatives of the equivalence classes under local unitaries and graph isomorphies of the connected graphs with $N=5,6,7$ vertices \cite{He03}.  The first [second] column depicts a representative of the class, that is the last [first] class of a given size $N$ to become PPT with respect to {\it some} partition. Hence the first [second] column contains those graphs, for which the PPT-criterion indicates, that the $N$--party distillable entanglement contained in these states might be most stable [instable] (among all graphs with the same number of vertices). Similarly the third [fourth] column shows a graph of the equivalence class, that is the last [first] of a given size $N$ to become PPT with respect to {\it all} partitions. Hence the third [fourth] column contains those graphs, for which the PPT-criterion indicates, that these states might be the last [first] to become $N$--party separable.}
\end{figure}
In this context we consider two graphs to belong to the same equivalence class if they can be transformed into each other by local unitaries {\it and} graph isomorphies. The latter corresponds to an exchange of particles, that maps neighboring particles onto neighboring particles.
We note that in this special case of noise due to the same individual depolarizing channel the notion of equivalence classes of graph states under local unitary 
transformations and graph isomorphies (i.e. particle exchange) is meaningful, since the decoherence process itself is invariant under these operations. As it can be seen in
 Fig.~\ref{fig:DepolLU} for connected graphs on $N=5,6,7$ the $N$-party GHZ states seem to be the first that looses $N$-party distillability.

Finally we will consider the case of individual dephasing channels $ {\cal D}\rho =   p \rho + \frac{1-p}{2} \left( \rho + \sigma_z \rho \sigma_z \right) $ (i.e. $p_0=\frac{1+p}{2},\, p_1=p_2=0,\, p_3=\frac{1-p}{2}$), for which the estimation (\ref{LambdaEstimation}) is no longer valid in general. It is straightforward to see that 
\be
q^{|U'|} \lambda_U \leq \lambda_{U+U'}\leq q^{-|U'|} \lambda_U
\ee  
holds for $q=\frac{1-p}{1+p}\leq 1$, since $\lambda_U = p^N q^{|U|}$. Similarly as in Eq.~(\ref{estim}), we therefore can bound the eigenvalues of the partial transpose 
$\lambda'_U$ with respect to the partition ${k}$ from above by
\be \lambda'_U \geq (1 + q^{|N_k|} + q - q^{-(|N_k|+1)}) \lambda_U\; .\ee
As depicted in Fig.~\ref{fig:DephasRing}, the above case of a ring ($|N_k|=2$) this inequality yields to the sufficient condition $q\geq 0.7549$ [$p \leq 0.1397$]
 for all one-versus-rest splits to have PPT and hence yields a stronger criteria for $N$-party distillable entanglement than Eq.~(\ref{upper1Pauli}).
\begin{figure}[th]
\includegraphics[width=8cm]{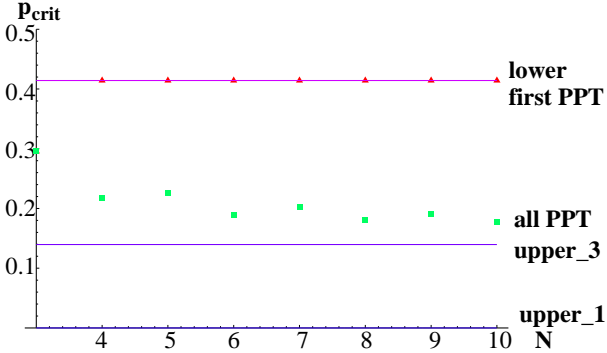}  
\caption{\label{fig:DephasRing} For the case of particles in rings of size $N\leq 10$, which individually decohere according to the same dephasing channel
 with parameter $p$:  the critical value $p_{\text{crit}}$, after which the first [last] partition becomes PPT $\triangle$ [$\Box$], the lower bound according 
to Sec.~\ref{lower} and the upper bounds according to Sec.~\ref{upper1} and Sec.~\ref{upper3}.} 
\end{figure} 
Note, that the lower bound actually coincides with the computed critical values for $p$, after which the ring first becomes PPT with respect to some partition.

\subsection{Generalization to weighted graph states}\label{weighted}

In this section we extend the previous results to a more general class of initial states, the so called weighted graph states. The graph states discussed so far 
arise from the Ising type interaction Hamiltonian $H_{kl} = |1\rangle^k\langle1|\otimes |1\rangle^l\langle1|$ (see Eq.(\ref{Ising})) acting on a collection of particles $V$ 
in the $\sigma_x$ eigenstate $|+\rangle^V$ for a fixed time $\varphi=\pi$ according to some interaction pattern specified by the graph.
 We will now allow the particles to interact according to the same Hamiltonian $H_{kl}$ but for different interaction times $\varphi_{kl}$. This corresponds to the situation of a disordered system as it occurs e.g. in a spin glass or semi--quantal Boltzmann gas. The interaction pattern
can similarly be summarized by a weighted graph, in which every edge is specified by a phase $\varphi_{kl}$ corresponding to the time the particles $k$ and $l$ have interacted.
 The  weighted graph state $|G\rangle$ is thus given by
\be
|G\rangle = \prod_{\{k,l\} \in E} U_{kl} |+\rangle^{V} \label{wgraph}
\ee
where the operations $U_{kl}$ are in this case given by
\be
U_{kl}\equiv e^{-i \frac{\varphi_{kl}}{4} \left(\mathbf{1}^{(k)}-\sigma_z^{(k)}\right)\otimes\left(\mathbf{1}^{(l)}-\sigma_z^{(l)}\right)}.
\ee  
In contrast to this straightforward extension of the interaction picture for weighted graph states, no such generalization of the stabilizer formalism 
(see Eq.~(\ref{stab})) in terms of generators within the Pauli group is possible. In particular this implies that the results of Sec.\ref{upper3} are no longer
 applicable to  weighted graph states.
But in the following we will show, that the other techniques established in the previous sections to obtain lower and upper bounds
 on the lifetime of entanglement can in fact be extended to cover also weighted graph states. Actually the following analysis will also hold for all states produced from $H_{kl}$ acting on an arbitrary product state $|\psi\rangle = |\psi_1\rangle \ldots |\psi_N\rangle$, which are not necessarily of the form $|\psi\rangle = |+\rangle^{\otimes V}$. Nevertheless, for the sake of simplicity we will restrict the following to this case.

\subsubsection{Lower bound on lifetime}

In order to obtain a lower bound on the lifetime of weighted graph states, we again provide an explicit protocol which allows the distillation of  maximally entangled states between all neighboring pairs of particles and thus to create arbitrary $N$--particle entangled states. In fact, we make use of the same protocol as in Sec.\ref{lower}, however the analysis turns out to be different. To be specific, we consider the state $\rho(t)$ which is obtained from a weighted graph state $|G\rangle$ subjected to decoherence --described by individual Pauli channels-- for time $t$.
We perform measurement in the eigenbasis of $\sigma_z$ on all but particles $k,l$ and determine the condition when the resulting reduced density operator
 $\rho_{kl}$ is distillable. We denote by $P_0^{j} =|0\rangle^j\langle 0|, P_1^{j} =|1\rangle^j\langle 1|$ projectors acting on particle $j$ and by ${\hat P}_0, {\hat P}_1$ the corresponding superoperators, i.e. ${\hat P}_0\rho =P_0 \rho P_0$. Similarly, we denote ${\hat U}_{kl}\varrho \equiv U_{kl}\varrho U_{kl}^{\dagger}$ the superoperator corresponding to the unitary operation $U_{kl}$. Note that the entanglement properties of the resulting state do not depend on the specific measurement outcomes. For notational convenience we restrict our analysis to the case, where the measurement result $0$ is obtained on all measured particles. Taking noise described by some Pauli channel 
$\mathcal{D}\rho=\sum_{i=0}^3 p_i \sigma_i \rho \sigma_i$, we thus have to consider the (unnormalized) state
\be
{\cal D}_{k}{\cal D}_{l}\prod_{j\not= k,l} {\hat P}_0^{j}{\cal D}_{j}  \prod_{\{a,b\} \in E} {\hat U}_{ab} |+\rangle^V\langle +|,\label{weightedDec}
\ee
Using that $P_0 \sigma_{0,3}=\sigma_{0,3} P_0$ and $P_0 \sigma_{1,2}=\sigma_{1,2} P_1$ we can rewrite ${\hat P}_0^{j}{\cal D}_{j} \varrho$ and obtain
\be
{\hat P}_0^{j}{\cal D}_{j}\varrho = \left({\cal M}^j_0 {\hat P}_0 + {\cal M}^j_1 {\hat P}_1\right) \varrho,
\ee
where
\bea
{\cal M}_0\varrho &=& p_0\, \varrho + p_3\, \sigma_3 \varrho \sigma_3, \nonumber\\
{\cal M}_1\varrho &=& p_1\, \sigma_1\varrho\sigma_1 + p_2\, \sigma_2 \varrho \sigma_2.
\eea
Choosing the computational basis $|U\rangle^{V\setminus\{k,l\}}=\sigma_x^U |0\rangle^{V\setminus\{k,l\}}$ on the measured particles, we thus can write
\be
\prod_{j\not= k,l}{\hat P}_0^{j}{\cal D}_{j} \varrho = \sum_{U \subseteq V\setminus\{k,l\} }\prod_{j \not=k,l} {\cal M}^{j}_{U_j} {\hat P}_{U_j}^{j} \varrho 
\ee
The projector commutes with the unitary operations $U_{ab}$ and we therefore obtain that Eq.~(\ref{weightedDec}) can be rewritten as
\bea
&&{\cal D}_{k}{\cal D}_{l} \sum_{ U \subseteq V\setminus\{k,l\}} \prod_{j \not= k,l} {\cal M}^{j}_{U_j}  \prod_{\{a,b\} \in E} {\hat U}_{ab}\nonumber \\
&&|++\rangle^{kl}\langle++| \otimes |U\rangle^{V\setminus\{k,l\}}\langle U | \; .
\eea
Note that $U_{ab}$ leaves $|U\rangle^{V\setminus\{k,l\}}$ invariant and it is thus sufficient to consider only $U_{ab}$ that act on particles $k,l$ and/or their neighbors, i.e. the set $I=N_k\cup N_l\cup k\cup l$. For all other particles $j\not\in I$ we thus have expressions of the form $\mathcal{M}_{U_j} |U_j\rangle^j\langle U_j| = |0\rangle^j\langle 0|$, i.e. these particles factor out. It follows that the reduced density operator $\rho_{kl}$ which is obtained by tracing out all but particles $k,l$ only depends on particles in the set $I$ but not on the other particles $j\in \bar{I}$ or errors (noise) effecting these other particles. This already shows that the lower bound on distillability for weighed graph states only depends on the (degree of the) corresponding interaction graph as well as the weights of the edges, but is independent of the size of the system $N$ as long as the degree of the graph itself does not depend on $N$. In particular, only the subgraph of particles $j \in I$ determines the entanglement properties of the reduced density operator $\rho_{kl}$. 

We have that $\rho_{kl}$ is given by
\be
&&\tr_{I\setminus\{k,l\}} \big( {\cal D}_{k}{\cal D}_{l} \sum_{ U \subseteq V\setminus\{k,l\}} \prod_{j \in I} {\cal M}^{j}_{U_j} 
 \prod_{a,b \in I \atop \{a,b\} \in E } {\hat U}_{ab}\nonumber \\
&&|++\rangle^{kl}\langle++| \otimes|U\rangle^{I\setminus\{k,l\}}\langle U| \big) \; ,
\ee
where the partial trace has to be performed for the remaining neighboring particles $I$ of $k$ and $l$ only. Thus the effect of noise can be localized to the region $I$ around the edge $\{k,l\}$ in question.
In principle, one can now obtain the explicit form of $\rho_{kl}$ for a given (weighted) graph and determine the condition for $p_i$ until when the reduced density operator $\rho_{kl}$ has non--positive partial transposition and remains thus distillable. The explicit formula is however rather complicated and not particularly illuminating.
 For the example of a depolarizing channel, it is clear that for smaller values of $\varphi_{kl}$ (i.e. a weaker edge between particles k and l) one obtains stronger threshold values on the parameter $p$ than given by Eq.~(\ref{lowerbDepol}), i.e. a shorter lifetime.

What is however more important in our context is that also for weighted graph states the lower bound on lifetime of distillable entanglement only depends on the (degree) of the corresponding interaction graph, but not on the size of the system $N$. Although the actual values of the lifetime will depend on the specific weights of the edges, for cluster--like and similar graph states there will be no scaling behavior with $N$. Moreover, in many cases such as rings, the edge with the smallest weight will give rise to the strongest threshold value condition  and will thus determine the lower bound on the lifetime of distillable $N$--party entanglement. Actually, it is sufficient if one can create maximally entangled pairs between pairs of particles in such a way that there exists a path between each pair of particles (i.e. entanglement between all pairs $\{k,l\}$ where the edges $\{k,l\}$ form a maximally connected graph). Thus the state is already $N$--party distillable, if the subgraph is connected, that is generated by all those edges, from which a Bell pair can be distilled. This implies that some edges in the original graph --even if they are very weak-- may not play a role if there exists another way to obtain singlets between all relevant pairs. For instance, if one considers a graph corresponding to a $1D$-cluster state, where each edge has weight $\pi$, and one adds in addition an edge $1,k$ with small weight $\varphi_{1k}$, then it is not necessary that entanglement between particles $1,k$ can be distilled (although the two particles are neighboring ones according to the graph $G$), but it is sufficient to distill entanglement between all pairs of particles $k,k+1$.

\subsubsection{Upper bound on lifetime}

The first method to obtain an upper bound to the lifetime of entanglement certainly also holds for arbitrary  weighted graph states,
 since it is independent of the initial state and reflects the time after which the decoherence process itself has become entanglement breaking. Conversely the upper bounds derived in this way cannot take into account whether the initial state is only slightly entangled or not. In Sec.~\ref{upper2} it turned out that for ordinary graph states the  upper bound is weaker than the first one derived in Sec.~\ref{upper1}. In contrast we will show in the following that the upper bound  presented in Sec.\ref{upper2} will give tighter upper bounds to the entanglement in weighted graph states, which contain vertices $k$ with only small interaction phases $\varphi_{kl}$ to all their neighbors $l\in N_k$. We use again the dynamical description of the weighted graph state $|G\rangle$ given by Eq.~(\ref{wgraph}). The influence of (phase) noise on the
 entanglement properties of $U_{kl}$ can be determined in a similar way as in Sec.\ref{upper2}, where we had a fixed angle $\varphi_{kl}=\pi$ for all $\{k,l\}\in E$.
 We remark that the upper bound obtained in Sec.\ref{upper2} for a general graph state is also valid for all graphs of the same kind where the edges are weighted.
 This is due to the fact that the operations $U_{kl}$ are most resistant to noise (i.e. remains entangling), if the angle is $\varphi_{kl}=\pi$, because in this case
 the operation is --in the ideal case-- capable to create maximally entangled states,
 while for all other values of $\varphi_{kl}$ only partially entangled states can be created. 

This observation immediately leads a way to obtain stronger upper bounds on the lifetime for weighted graph states. To this aim, one determines the threshold values
 $p_z^{kl}$ when the state $\tilde \rho(t)$ (Eq.~(\ref{tilderho})) becomes separable. The value of $p_z^{kl}$ now depends on $\varphi_{kl}$. One finds that the corresponding density operator $\tilde D_{k_1k_2l_1l_2}((p^{kl}_z)^{1/|N_k|},(p^{kl}_z)^{1/|N_l|})$ in Eq.~(\ref{Dzstate}) again has support only in a four dimensional subspace and is given by 
\be
\tilde D_{k_1k_2l_1l_2} = \sum_{i,j=0}^1 \lambda_{ij} |\Phi_{ij}\rangle\langle\Phi_{ij}|,
\ee
with 
\bea 
\lambda_{00} & = &\frac{1}{4} \left(1+(p^{kl}_z)^{\frac{1}{|N_k|}}\right)\left(1+(p^{kl}_z)^{\frac{1}{|N_l|}}\right) \cr
\lambda_{01} & = &\frac{1}{4} \left(1+(p^{kl}_z)^{\frac{1}{|N_k|}}\right)\left(1-(p^{kl}_z)^{\frac{1}{|N_l|}}\right) \cr
\lambda_{10} & = &\frac{1}{4} \left(1-(p^{kl}_z)^{\frac{1}{|N_k|}}\right)\left(1+(p^{kl}_z)^{\frac{1}{|N_l|}}\right) \cr
\lambda_{11} & = &\frac{1}{4} \left(1-(p^{kl}_z)^{\frac{1}{|N_k|}}\right)\left(1-(p^{kl}_z)^{\frac{1}{|N_l|}}\right)  \; .
\eea 
The orthogonal states  $|\Phi_{ij}\rangle$ are given by 
$(\sigma_z^{(k_2)})^i\otimes (\sigma_z^{(l_2)})^j |\Phi_{00}\rangle$, where
\be
|\Phi_{00}\rangle \equiv |\tilde 0\rangle_k |\tilde 0\rangle_l + |\tilde 0\rangle_k |\tilde 1\rangle_l + |\tilde 1\rangle_k | \tilde 0\rangle_l + e^{i\varphi_{kl}} |\tilde 1\rangle_k |\tilde 1\rangle_l
\ee
with $|\tilde 0\rangle_{k}=|00\rangle_{k_1k_2}, |\tilde 1\rangle_{k}=|11\rangle$ and similar for particles $l_1l_2$.
 We have again, that $\tilde D_{k_1k_2l_1l_2}$ is separable if and only if the partial transposition is positive, which leads to the threshold value $p_z^{kl}$.
\begin{figure}[th]
\includegraphics[width=8.5cm]{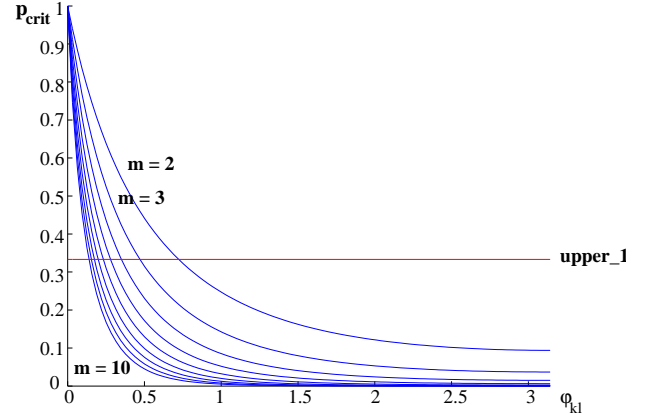}  
\caption{\label{fig:weightedupper} For the case of individual depolarizing channels with parameter $p$ the threshold value $p_{crit}=\frac{p_>}{2+p_>}$ (see Eq.~(\ref{Depolp_z})) as a function of the interaction phase $\varphi_{kl}\in [0,\pi]$ for edges $\{k,l\}$ between two vertices $k$ and $l$ with the same increasing degree $m=|N_k|=|N_l|=2,3,\ldots,10$. The horizontal line depicts the upper bound according to Sec.~\ref{upper1} in this case.}
\end{figure}
The separability of the weighted graph state can then be determined in a similar way as in Sec.\ref{upper2}. In order to make the operation $U_{kl}$ separable,
 $\sigma_z$ noise with $(p_z^{kl})^{1/{|N_k|}}$ and $(p_z^{kl})^{1/{|N_l|}}$  is required at vertices $k$ and $l$. At a given vertex $k_0$, this leads to a required total value of
 $p_>^{k_0} = \min_{l \in N_{k_0}} p_>^{k_0l}$ such that all operations $U_{k_0l}$ become separable. The threshold value $p_>$, below which the state is fully separable,
 is finally obtained by taking the minimum over all $p_>^{k}$ (that is over all vertices). That is
\be\label{upperWei}
p_> \equiv \min_k p_>^{k},
\ee
and the state $\rho$ is certainly separable for $p_z < p_>$. For the different decoherence models $\mathcal{E}$ allowing for an extraction of a 
dephasing component (see Eq.~(\ref{Decomp})) one finally has to insert the relation between the dephasing parameter $p_z$ and the noise parameter of $\mathcal{E}$ in order
 to obtain the announced upper bound on the lifetime for weighted graph states. Fig.~\ref{fig:weightedupper} depicts the critical value for the depolarizing parameter $p$ in Eq.~(\ref{Depol}) as a function of the weight $\varphi_{kl}$ at the edge in question. As it was already mentioned in Sec.~\ref{upper2}, for a fixed phase $\varphi_{kl}$ the obtained upper bound on the lifetime decreases with the degree of the neighboring particles $k$ and $l$ (in contrast to the corresponding behavior of the lower bound in Sec.~\ref{lower}). Moreover Fig.~\ref{fig:weightedupper} shows, that for any degree $m$ there always exists a range of values for $\varphi_{kl}$, for which the analysis of this Sec. provides a stronger upper bound $p_{\text{crit}}$ than the condition derived in Sec.~\ref{upper1}.

\section{Block wise entanglement and re--scaling} \label{blockwise}

Entanglement is a concept which can only be defined between subsystems of the whole system. In our previous analysis, we have identified subsystems with parties, i.e. we have investigated the lifetime of true $N$--party entanglement. One can however also consider a slightly more general concept, where subsystems are formed by a collection of several parties (see Sec.~\ref{Entanglement}). Also in this case one can investigate entanglement properties between $M$ such subsystems. That is, one can consider a partitioning of the $N$--party system into $M \leq N$ groups and investigate the (distillable) entanglement between these $M$ groups, where each of these groups consists of one or several of the initial parties. Whenever $M<N$, one can have that the state is still $M$--party entangled although it contains no longer $N$ party entanglement. Considering such coarser partitions allows one to investigate the change of the kind of entanglement in time and to determine an ``effective size'' of the entanglement present in the system. One can determine for each kind of entanglement the corresponding lifetime.

\subsection{Block wise entanglement: Distillability and lifetime}\label{Def_blockwise}

Given an $N$ party system, we consider a partitioning of the $N$ parties into $M<N$ groups ($M$--partitioning). Parties within a given group are allowed to perform joint operations are considered as a single subsystem with a higher dimensional state space. We are interested in the entanglement between these $M$ subsystems. A density operator $\rho$ is called $M$--party distillable with respect to a certain $M$--partitioning if from (asymptotically) many copies of $\rho$ one can create some irreducible entangled pure state by means of {\em local} (in the sense of the $M$--partitioning in question) operations and classical communication. Similarly, a density operator $\rho$ is separable with respect to a certain $M$--partitioning if it can be written as convex combination of products states (in the sense of the $M$--partitioning in question).

\begin{figure}[th]
\includegraphics[width=8cm]{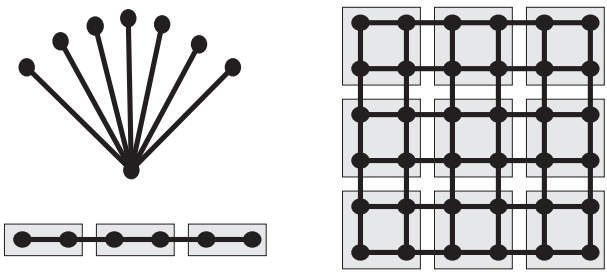}  
\caption{\label{fig:rescale} For blockwise entanglement different partitionings of particles into groups are considered.}
\end{figure} 

We say that a density operator is $M$--party distillable if it is distillable with respect to at least one $M$--partitioning. It is obvious that if $\rho$ is $M$--party distillable, it as also $M'$--party distillable for $M' \leq M$. The maximal possible $M$ such that a density operator $\rho$ is $M$--party distillable can be interpreted as a measure of the ``size'' of entanglement, as it provides the maximal number of subsystems (groups, blocks) which are (distillable) entangled. We will investigate the lifetime of $M$--party distillable entanglement for all $M$. We remark that $M$--partitionings might be completely different in nature. Consider for instance a system consisting of 90 parties $A_k, k=1,\ldots ,90$. Possible $3$--partitionings include for instance $(A_1)-(A_2)-(A_3,A_4,\ldots, A_{90})$ as well as $(A_1,\ldots, A_{30})-(A_{31},\ldots, A_{60})-(A_{61},\ldots, A_{90})$. For a detailed description of $M$--partitionings as well as necessary and sufficient conditions for $M$--party distillability we refer the reader to Ref. \cite{Du00}.

Equivalently one may say that we are considering a coarse-graining of the partition and investigate the entanglement properties under coarse graining. As a particular instance of such coarse graining we will investigate partitionings that correspond to re--scaling of the size of the subsystem as it is used in statistical mechanics. Consider for instance $N$ particles which are arranged on a regular rectangular (two--dimensional) lattice. The finest partition corresponds to considering each particle individually as a single subsystem. Coarsening of the partition may e.g. take place by considering blocks of $n\times n$ particles (arranged as a square) as a single subsystems. That is, for $n=2$ one considers a specific $M=N/4$ partitioning, while for arbitrary $n$ we have $M=N/n^2$ groups of parties/subsystems, each formed by $n^2$ particles. This concept is also illustrated in Fig.~\ref{fig:rescale}. We will investigate how (distillable) entanglement changes under such re--scaling and determine asymptotic properties of the lifetime of $M$--party entanglement for macroscopic number of particles, $N\to \infty$.

\subsection{Lifetime of GHZ states under coarse graining}\label{block}

\subsubsection{Lifetime of entanglement in large $T$ limit of reservoir}\label{blockGHZ}

As in Sec.~\ref{GHZ}, we consider the lifetime of distillable entanglement for GHZ states when each particle is individually coupled to a thermal reservoir with $T\to \infty$, described by a depolarizing quantum channel. In order to determine the lifetime of $M$--party entanglement, we can make use of the results obtained in Sec.~\ref{GHZ}, together with the classification of GHZ--diagonal states of Ref. \cite{Du00}. Recall that the partial transposition with respect to a group (subsystem) $B_k$ which contains exactly $k$ parties is positive, $\rho(t)^{T_{B_{k}}} \geq 0$ if and only if $p^N \leq 2\lambda_k$ (see Eqs. (\ref{lambdakGHZ},\ref{cond})). In addition, we have $\lambda_1 \geq \lambda_2 \geq \ldots \geq \lambda_{[N/2]}$ (see Eq.~(\ref{condlambda})). This implies that the size of the subsystem, i.e. the number of particles that are contained in a subsystem, determine when the corresponding partial transposition becomes positive. We have that the partial transposition with respect to a single party, $\rho(t)^{T_{A_{k}}}$, is the first one that becomes positive, while the partial transposition with respect to a larger subsystem is more stable, i.e. becomes positive at a later time. 

For a given $M$--partitioning, above observation immediately allows one to identify the bipartite partition (which contains the $M$--partition in question) which determines the lifetime of distillable $M$-party entanglement. In particular, the partial transposition with respect the subsystem that contains the {\em smallest} number of parties is the first one to become positive. We use that a necessary --and in the case of states of the form $\rho(t)$ we deal with in this case also sufficient-- condition for $M$--party distillability is that the partial transposition with respect to all subsystems $B_k$ forming the $M$--partitioning is non--positive. To be exact, one also needs that all partial transposition with respect to groups formed by several such subsystems $B_k$ are also non--positive, which is in our case however automatically fulfilled as partial transpositions with respect to larger groups are more stable than with respect to smaller groups. If only one of all these partial transpositions with respect to various subsystems is positive, it follows that the state $\rho(t)$ is no longer $M$--party distillable (see Ref. \cite{Du00}). Thus we have that the lifetime of $M$--party distillable entanglement with respect to a given $M$--partitioning is determined by the size of the smallest subsystem of the corresponding $M$--partitioning. For instance, if one considers an arbitrary $M$--partitioning ($M<N$) that contains as one subsystem a single party, say $A_1$, we have that the lifetime of entanglement with respect to this partitioning is completely determined by the partial transposition with respect to party $A_1$, i.e the condition $p^N \leq 2 \lambda_1$. In particular, the lifetime of distillable $M$--party entanglement with respect to any such partitioning is exactly the same as for $N$--party entanglement (corresponding to a $N$--partitioning where each subsystem contains a single party). 

It follows that for a given $M$, $M$--party entanglement with respect to a specific $M$--partitioning has longest lifetime if all groups have (approximately) the same size. For a minimal group size of $m$ particles, a $N$--particle GHZ state can contain at most $M\equiv [N/m]$ such groups of size $m$. This allows one to obtain the maximum lifetime of $M$--party entanglement which is determined by $p^N \leq 2 \lambda_{m}$ (Eq.~(\ref{cond}) with $k=m$).

Determining the threshold value $p_{\rm crit}$ involves the solution of a polynomial equation of degree $N$, which can be done numerically in an efficient way. One can however also determine analytic lower and upper bounds on the lifetime of $M$--party entanglement. One obtains an {\em upper bound} on the lifetime of $M$--party entanglement if one approximates $\lambda_m$ by some $\tilde\lambda_m \leq \lambda_m$ and investigate the condition 
\be
p^N \leq 2 \tilde \lambda_m,
\label{condtildelambda}
\ee
as in this case automatically also $p^N \leq 2 \lambda_k$ and thus the partial transposition with respect to a group that contains $m$ parties is certainly positive. We can e.g. choose
\be
\tilde\lambda_m \equiv (1-p)^m(1+p)^{N-m}/2^{N+1},
\ee
which obviously fulfills $\tilde \lambda_k \leq \lambda_k$. The condition Eq.~(\ref{condtildelambda}) can then be rewritten as
\be
m \leq N \frac{\log[2p/(1+p)]}{\log[(1-p)/(1+p)]}.\label{condm1}
\ee
Recall that in the (optimal) case where all subsystems have the same size we have $M\equiv [N/m]$. Thus we find that a $N$--party system is certainly no longer $M$--party entangled if
\be
M \geq \frac{\log(1-p)-\log(1+p)}{\log(2p)-\log(1+p)}.\label{condM}
\ee

Eq.~(\ref{condM}) is a central result in our analysis of the properties of the lifetime for GHZ states under re--scaling. We have illustrated Eq.~(\ref{condM}) in Fig.~\ref{fig:upperGHZ1} and \ref{fig:upperGHZ2}. 

On the one hand, Eq.~(\ref{condM}) provides an upper bound on the lifetime $\kappa \tau_M$ of $M$--party entanglement in the system. This upper bound can be obtained by determining $\kappa \tau_M$ for a fixed $M$ from Eq.~(\ref{condM}), where $\kappa t$ is again given by $p=e^{-\kappa t}$.
On the other hand, for a fixed time $t$ one can determine from  Eq.~(\ref{condM}) the maximum $M$ of distillable multipartite entanglement in the system. That is, the effective size of entanglement after a time $t$ can be obtained this way. One observes (see Fig.~\ref{fig:upperGHZ1} and \ref{fig:upperGHZ2}) that the maximum $M$ rapidly decreases with $t$. For small times, i.e. $\kappa t \ll 1$, one finds that $M$ scales as
\be
M \approx -2\log(\kappa t)/(\kappa t).
\ee
For $\kappa t> 0.8049$ (which is obtained from Eq.~(\ref{condM}) by setting $M=2$), we have that also $2$--party entanglement disappears. In fact, the state becomes fully separable as all partial transposes are positive (which is a sufficient condition for separability for such states \cite{Du00}). We emphasize that the upper bound on the lifetime of $M$--party entanglement is {\em independent} of the number of particles $N$, in particular it is also valid for a macroscopic number of particles and even for $N\to \infty$.

On the one hand this implies that even if the size of the groups $m=N/M$ goes to $\infty$ (for a total number of particles $N \to \infty)$, the maximum number of groups $M$ that can remain entangled after a time $t$ is {\em finite}, i.e. the maximum effective size of entanglement is limited for any time $t$. 
On the other hand it follows that in the limit $N \to \infty$ any partitioning in groups with finite size $m$ leads to a vanishing lifetime of the corresponding $M=N/m$ party entanglement (see Eq.~(\ref{condm1})). Only if one considers the limit where the number of subsystems $M$ is fixed as $N \to \infty$, i.e. the size of each of the groups itself is macroscopic and $m\to \infty$, one obtains that the lifetime of the corresponding $M$--party entanglement is {\em finite}. 
We remark that above results also enable one to obtain (upper bounds) on the lifetime of $N$--party entanglement by considering $m=1$ and $M=N$ (see also Sec.\ref{GHZ}).

In an analogous way one can derive a {\em lower bound} on the lifetime of $M$--party entanglement. To this aim, one uses that if the partial transposition with respect to the smallest subsystem is still non--positive, also all other relevant partial transpositions of the system with respect to all other subsystems (and combinations thereof) are non--positive. For states of the form Eq.~(\ref{rhot}) this ensures that the state $\rho(t)$ is $M$--party distillable \cite{Du00}. Thus we can derive an analytic lower bound on the lifetime of $M$--party distillability by considering the condition Eq.~(\ref{cond}) with $k=m$ and upper bound $\lambda_m$ by some $\tilde \lambda'_m$. We have that if $p^N > 2\tilde \lambda'_m \geq 2 \lambda_m$, then $\rho(t)$ is certainly $M$--party distillable with respect to a partition which consists of $M$ subsystems of size $m=[N/M]$. 
We choose
\be
\tilde\lambda'_m \equiv 2(1-p)^m(1+p)^{N-m}/2^{N+1},
\ee
which can readily be checked to fulfill $\tilde \lambda'_m \geq \lambda_m$. The condition $p^N > 2\tilde \lambda'_m$ can be rewritten and one finds that for
\be
M \leq \frac{\log[2(1-p)]-\log(1+p)}{\log(2p)-\log(1+p)}
\ee
all partial transposition with respect to this $M$--partitioning are certainly non--positive, which already ensures that the state $\rho(t)$ is $M$--party distillable.

At first sight, our results seem to contradict the ones obtained by Simon and Kempe \cite{Si02}. They observed that the threshold value for $p$ when considering the partition $(N/2)-(N/2)$ decreases with the size of the system $N$. Based on this observation, they conclude that GHZ states of more particles are more stable against local decoherence. However, as pointed out in the discussion above, the effective number of subsystem that remain entangled {\em decreases} with time. The entanglement becomes bipartite when approaching the threshold value found by Simon and Kempe. In fact, the lifetime of genuine (distillable) $N$--party entanglement decreases with the size of the system $N$.

\subsubsection{Quantum optical channel}

One can now perform a similar analysis of the lifetime of $M$--party entanglement for more general couplings of the individual particles to the environment described by a general quantum optical master equation. We have already determined the condition such that the partial transposition with respect to a group of $k$ parties is positive in Sec.\ref{GHZ_QO} (see Eq.~(\ref{condN})). Following the line of argumentation if Sec.\ref{blockGHZ} and using the notation of Sec.\ref{GHZ_QO}, it is straightforward to obtain an upper bound on the lifetime of $M$--party entanglement. In particular, one has that positivity of at least one of the partial transpositions with respect to a specific subsystem ensures that $\rho(t)$ is no longer $M$--party distillable. That is, when considering (as in the previous section) a $M$--partitioning of the $N$--party system into $M$ subsystems, each of size $m=N/M$, an upper bound on the lifetime is given by Eq.~(\ref{condN}) with $k=m$. In fact, also in this case we have that the subsystems that contains the smallest number of parties gives rise to the strongest condition on the lifetime of $M$--party entanglement, i.e. the corresponding partial transposition is the first one to become positive. This can be seen by considering the condition for positivity of the partial transposition with respect to a group of $k$ particles given by Eq.~(\ref{condPPTg}), $[b^N/2]^2 \leq \lambda_k \lambda_{N-k}$. We have that for $1 \leq k \leq [N/2]$ 
\be
\lambda_k\lambda_{N-k} \geq \lambda_{k+1}\lambda_{N-k-1},\label{prodlam}
\ee
(which can be checked by direct computation and considering separately the cases $(ac)>0$ and $(ac) \leq 0$). From this observation the claim already follows, as Eq.~(\ref{prodlam}) ensures that if $\rho(t)$ has positive partial transposition with respect to $k+1$ parties, it automatically has also positive partial transposition with respect to $k$ parties. Hence the subsystem with the smallest number of parties determines the lifetime of $M$--party entanglement. Using again that for a fixed $M$ one obtains the longest lifetime if all $M$ groups have the same size $m=[N/M]$, it follows from Eq.~(\ref{condN}) that $\rho(t)$ is certainly no longer $M$--party distillable entangled if 
\be
M \geq  \frac{\log(ac)-\log[(1-a)(1-c)]}{\log(ac)+B t},
\ee
where $a,b,c$ are defined by Eq.~(\ref{abc}).
Thus an analogous discussion as in Sec.\ref{blockGHZ} applies, i.e. the scaling properties of the lifetime of GHZ states with the size of the system $N$ and the number of subsystems $M$ is {\it similar} for different couplings of the particles to the environment. Only in singular cases (such as zero temperature corresponding to $s=0$ or $s=1$), one discovers a different behavior. \\

Considering general graph states instead of GHZ states a detailed analysis of their blockwise entanglement has not been accomplished yet. Nevertheless, the scaling behavior of $M$-party entanglement  is restricted to a range between the upper and lower bounds of Sec.~\ref{graph}, which in the case of cluster and similar graph states were shown to be independent of the number $N$ of particles. In this sense also the scaling behavior of blockwise entanglement in these states must be essentially independent of the size of the system.


\subsection{Lifetime of encoded entangled states}\label{encoded}

Until now we have seen two different kind of scaling behaviors with respect to the number of particles $N$. For GHZ states, we found that the lifetime of distillable entanglement {\em decreases} with $N$, while for cluster states (and similar graph states) we have that the lifetime is {\em independent} of $N$. In this Section we will show that certain states can show a different scaling behavior, namely that the lifetime of (block--wise) entanglement can even {\em increase} with the number of particles $N$. Examples of such states are provided by encoded entangled states, i.e. entangled states which are formed by logical qubits where each of the logical qubits corresponds to the codewords of a (concatenated) quantum error correction code. We find that entanglement between $M$ such logical qubits is maintained. Each of the logical qubits forms a subsystem of size $m$, and we consider entanglement between $M$ such subsystems. We will show that as $m$ increases (but is still of finite size), there exist states such that the maximum number of subsystems $M$ that remain entangled increases. In addition, for a fixed $M$ we have that the lifetime of $M$--party entanglement increases as the block--size $m$ increases and tends to infinity as $m \rightarrow \infty$. This shows that (encoded) macroscopic entangled states --even of GHZ--type-- can persist for long times. Note that this behavior is in contrast to non--encoded GHZ states, were on the one hand the maximum number of subsystems $M$ that remain entangled after a certain time is finite and on the other hand also the lifetime of $M$ party entanglement is finite (even as $m \rightarrow \infty$). 

One can interpret these results in the sense that the time in the encoded system is slowed down as compared to the time in the original system. This provides an alternative view on quantum error correction and allows one to understand why encoded macroscopic superposition states can be produced and maintained on a quantum computer.

\subsubsection{Quantum error correcting codes}

We consider two orthogonal states of $m$ qubits, $|0_L\rangle, |1_L\rangle \in (\C^2)^{\otimes m}$ which correspond to codewords of some error correcting code and constitute the basis state of a ``logical qubit''.
In the following we will consider an optimal error correcting code which allows to correct an arbitrary error on one of the particles and uses five physical qubits to encode one logical qubit, e.g. the five qubit Steane code \cite{St98}. In the following discussion we will assume that each of the physical qubits is coupled to an independent environment and the the individual coupling is described by depolarizing quantum channels ${\cal D}(p)$ (corresponding to the coupling to a heat bath in the large $T$ limit), Eq.~(\ref{Depol}), with $p=e^{-\kappa t}$. We consider the evolution of an arbitrary state of a single logical qubit
\be
|\phi_L\rangle = \alpha |0_L\rangle +\beta|1_L\rangle,
\ee
under the influence of depolarizing channels acting on each of the individual particles, i.e. 
\be
\varrho(t)={\cal D}_1(p){\cal D}_2(p){\cal D}_3(p){\cal D}_4(p){\cal D}_5(p) |\phi_L\rangle\langle\phi_L|.
\ee
The action of the map ${\cal D}_k$ on physical qubit $k$ is such that with probability 
\be
q \equiv (3p+1)/4,
\ee 
no error occurred, while with probability $(1-q)$ the qubit was effected by some error. In particular, we have that one of the three possible errors described by $\sigma_k$, $k=1,2,3$ occurred with probability $(1-q)/3=(1-p)/4$. Considering now the logical qubit consisting of five physical qubits, we know that there exists a sequence of operations (error syndrome measurement followed by a correction step depending of measurement outcome) such that the state of the five qubits remains in the subspace spanned by $\{|0_L\rangle,|1_L\rangle\}$ and the logical qubit remains in the initial state $|\phi_L\rangle$ as long as no or only a single error in one of the physical qubits occurred. That is, with probability 
\be
q_L=q^5 + 5 q^4(1-q),
\ee
no or only a single (correctable) error happened, while with probability $(1-q_L)$ the logical qubit was effected by some error. By applying (correlated) random unitary operations on the subspace spanned by $\{|0_L\rangle,|1_L\rangle\}$ at $t=0$ and $t$, one can achieve that the errors can again be described by {\em white noise} acting on the logical qubit, i.e. a map of the form
\be
{\cal D}_L (p_L) \rho_L = p_L \rho_L + \frac{1-p_L}{4} \sum_{k=0}^3 \sigma_k^{(L)} \rho_L \sigma_k^{(L)},
\ee
where $p_L=(4q_L-1)/3$ and $\sigma_k^{(L)}$ denote Pauli operators acting on the logical qubits, e.g. $\sigma_1^{(L)}=|0_L\rangle\langle 1_L|+|1_L\rangle\langle 0_L|$. The parameter $p_L$ is related to the initial $p$ via
\be
p_L=(3p+1)^4(4-3p)/192-1/3.\label{pL}
\ee
That is, the action of the decoherence process on the logical qubit can (after performing a correction step plus depolarization) be described by a depolarizing channel acting on the logical qubit, where the parameter $p_L$ can be obtained from $p$ --the parameter describing the decoherence process of the individual physical qubits-- by Eq.~(\ref{pL}). We thus have that a logical qubit where each of the particles is subjected to decoherence for time $\kappa t = -\log(p)$ behaves as if it was subjected to decoherence described by a depolarizing channel acting on the logical qubit for time $\kappa t_L = -\log(p_L)$. We remark that if $p>0.82517$ ($\kappa t< 0.1921658$), we have that $p_L > p$ ($\kappa t_L < \kappa t$), that is the logical qubit is less effected by decoherence as a physical one. In other words, the effective time $t_L$ for which the decoherence acts on the logical qubit is smaller than the physical time and thus the decoherence process on the logical qubit is slowed down.

In a similar way, one can consider logical qubits which are formed by codewords of concatenated quantum error correction codes with $k$ concatenation levels. A logical qubit consists in this case of $5^k$ physical qubits. This follows from the fact that at concatenation level $j$, each logical qubit at level $j-1$ is replaced by five such logical qubits (of level $j-1$) which form the new logical qubit at concatenation level $j$. Following the same reasoning as in the case of a single concatenation level, one obtains that a logical qubit at concatenation level $j$ is subjected to decoherence described by a depolarizing quantum channel acting at the logical qubit, where the noise parameter $q_j$ is related to the noise parameter $q_{j-1}$ of the (logical) qubits at concatenation level $j-1$ by
\be
q_j=q_{j-1}^5 + 5 q_{j-1}^4(1-q_{j-1}),\label{qj}
\ee
where $q_j = (3p_j+1)/4$ and $q_0=q$ ($p_0=p$) corresponds to decoherence acting on the physical qubits. That is, at each concatenation level the effective time $t_j$ for which the decoherence acts on the logical qubit of level $j$ decreases as compared to the physical time $t \equiv t_0$ as long as $\kappa t < 0.1921658$. For $\kappa t_0 \equiv 4/3 \epsilon_0 \ll 1$ we have that
\be
\epsilon_{j} \approx 10 \epsilon_{j-1}^2,
\ee
from which follows
\be
\kappa t_j \approx \frac {(7.5 \kappa t)^{2^j}}{7.5}. \label{tj}
\ee
That is, the effective time $\kappa t_j$ for the decoherence acting on the logical qubit is drastically decreased. For instance, if $\kappa t_0 = \kappa t = 0.01$, we have that $\kappa t_1 \approx 7.5 \times 10^{-4}, \kappa t_2 \approx 4.22 \times 10^{-6}, \kappa t_3 \approx 1.33 \times 10^{-10}, \kappa t_4 \approx 1.34 \times 10^{-19}, \kappa t_5 \approx 1.34 \times 10^{-37}, \kappa t_6 \approx 1.35 \times 10^{-73}$. Recall that the number of physical qubits that form a logical qubit at concatenation level $j$ is given by $5^j$, which corresponds to $5,25,125,625,3125,15625$ respectively. In particular, this implies that by using $625$ particles to encode one qubit (i.e. $j=4$), the effective time $t_4$ as compared to the physical time $t$ can be decreased by a factor of $10^{17}$. 

\subsubsection{Lifetime of block--wise entanglement of encoded states}

Using the effective change in the timescale for logical qubits, it is now straightforward to determine lifetime of encoded entangled states. We consider block--wise entanglement, i.e. entanglement between $M$ subsystems where each subsystem consists of $m$ qubits. We consider lifetime of different kinds of entangled states which are formed by logical qubits and where each physical qubit is subjected to decoherence described by a depolarizing quantum channel (Eq.~(\ref{Depol})) as in Sec.\ref{largeT}. We emphasize that a logical qubit behaves in this context in exactly the same way as a physical qubit, the only difference is that the effective time $t_L$ (or noise parameter $p_L$) is different. This implies that the results obtained for GHZ and graph states can be directly used to obtain lifetime of such encoded entangled states.

Consider an entangled state of GHZ--type which is given by
\be
|GHZ_L\rangle = \frac{1}{2}(|0_L\rangle^{\otimes M} + |1_L\rangle^{\otimes M}),\label{GHZL}
\ee
where $\{|0_L\rangle,|1_L\rangle\}$ are codewords of (concatenated) quantum codes (with $j$ concatenation levels) formed by $m$ physical qubits which represent a logical qubit. When considering block--wise entanglement between $M$ blocks of $m=5^j$ qubits, this state behaves in exactly the same way as a $M$ particle GHZ states (Eq.~(\ref{GHZstate})) consisting of physical qubits where one considers blocks of size $1$. The only difference is that the effective timescale $t_j$ (effective noise parameter $p_j$) is changed according to Eqs.~(\ref{qj},\ref{tj}). 

When considering the original GHZ state and $\kappa t = 10^{-2}$, we obtain from Eq.~(\ref{condM}) (i) After a time $\kappa t =10^{-2}$, the maximum number of blocks that can be entangled is upper bounded by $M=1057$; (i) For $M=1057$, $\kappa t = 10^{-2}$ provides an upper bound on the lifetime of $M$ blocks (where in both cases each block may have arbitrary size, i.e. $m\to \infty$). 

For an encoded GHZ state (Eq.~(\ref{GHZL}) we have the following results: (i) The maximum number of blocks of size $m=5^j$ (i.e. logical qubits) that is entangled after a time $\kappa t=10^{-2}$ is determined by Eq.~(\ref{condM}), where $p$ in this equation is given by $e^{-\kappa t_j}$ and $\kappa t_j$ is the effective time. We find that for $j=1, M=2.103 \times 10^{4}; j=2, M=6.195 \times 10^{6}; j=3, M=3.510 \times 10^{11}$, i.e. the number of blocks of fixed size $m=5^j$ that remain entangled after some time $\kappa t$ is drastically increased. (ii) An upper bound on the lifetime of $M=1057$ blocks consisting of logical qubits of size $m=5^j$ is provided by $\kappa t_j = 10^{-2}$, where $\kappa t_j$ is the effective time. One can determine the time $\kappa t$ (which provides an upper bound on the lifetime of such systems) using the recursive formula Eq.~(\ref{qj}). One finds that for $j=1, \kappa t = 0.0382$; $j=2, \kappa t=0.0778$;  $j=3, \kappa t=0.1149$; $j=4, \kappa t=0.1431$; $j=5, \kappa t=0.1621$. Again, one sees that the lifetime of $M$--party entanglement is increased as compared to the original GHZ state, although here we considered only blocks of finite size $m=5^j$, while we allowed for blocks of arbitrary size $m \to \infty$ in the case of (original) GHZ state. We remark that one can only expect that the encoded system has longer lifetime as compared to the original GHZ state as long as $\kappa t< 0.1921658$, since only in this case $t_j < t$. 

In a similar way, the lifetime of $M$--party cluster- and graph states (formed by logical qubits) is enhanced when considering such states formed by logical qubits.


\section{Summary and conclusions}\label{summary}

In this paper we have investigated the lifetime of (distillable) entanglement under the influence of decoherence. We found that the qualitative behavior of different kinds of entangled states are largely independent of the specific decoherence model. In particular, we found for (essentially) all decoherence models with individual coupling of particles to (independent) environments that the lifetime of GHZ states decreases with the size of the system. On the other hand the lifetime of cluster states and graph states with a constant degree (which does not depend on $N$) is independent of the number of particles $N$. The last observation can even be extended to all decoherence models which correspond to some correlated but localized noise, i.e. where the Kraus operators of the corresponding map act only non-trivially on a finite, localized number of subsystems. We have also considered lifetime of entanglement between subsystems of different size, which allowed us to determine the scaling behavior of entanglement under re--scaling of the size of the subsystems. While for cluster states there is essentially no change in the scaling behavior with $N$, for GHZ states we found that (i) the lifetime of block--wise entanglement for any number of blocks that contain only a finite number of particles $m$ tends to zero as $N \to \infty$, while it can become finite if the blocks itself become macroscopic, i.e. $m \to \infty$ as $N \to \infty$;  (ii) the number of blocks $M$ that remain entangled after a certain time $t$ is finite, independent of the block size $m$. In addition, we have shown that for encoded entangled states the number of blocks that can be entangled after a certain time can be drastically increased and the lifetime of $M$--party entanglement can be enhanced.

Our results suggest a remarkable robustness of certain kinds of macroscopic entangled states --namely all graph states with constant degree-- under various kinds of decoherence.


This work was supported by the \"Osterreichische Akademie der Wissenschaften through project APART (W.D.), the European Union (IST-2001-38877,-39227) and the Deutsche Forschungsgemeinschaft.

\section*{Appendix A}
In this appendix we will give a more detailed analysis of the statement $(ii)$  of Sec.\ref{upper2}.
Given an arbitrary channel 
\be \mathcal{E}\rho = \sum_{i,j=0,1,2,3} \, p_{ij} \, \sigma_i \rho \sigma_j \; ,\ee
we first consider the question, whether it is possible to decompose it into a dephasing channel $\mathcal{D}\rho = \frac{1+p_z}{2} \rho + \frac{1-p_z}{2} \sigma_z \rho \sigma_z$
with $p_z\in\left[0,1\right]$ followed by some arbitrary noise channel $\mathcal{E'}\rho = \sum_{i,j=0,1,2,3} \, q_{ij} \, \sigma_i \rho \sigma_j$, i.e.
 \be \mathcal{E} = \mathcal{E'}\circ \mathcal{D}\; . \label{Decomp} \ee
Since the upper bound derived in Sec.\ref{upper2} becomes tighter with an decreasing dephasing parameter $p_z$, we will also try to minimize $p_z$ for those channels 
$\mathcal{E}$, for which an extraction of a dephasing part is non trivial, i.e. $p_z<1$.
First of all, with the matrices $\mathbf{P}=(p_{ij})$, $\mathbf{Q}=(q_{ij})$ and
 \be \mathbf{M} = \left( \begin{array}{cccc} 0 & 0 & 0 & 1 \\  0 & 0 & i & 0 \\ 0 & -i & 0 & 0 \\ 1 & 0 & 0 & 0 \\\end{array} \right) = \mathbf{M}^\dagger \ee
 we can rewrite Eq.~(\ref{Decomp}) as a matrix equation
\be \mathbf{P} = \frac{1+p_z}{2} \mathbf{Q} \,+\, \frac{1-p_z}{2} \mathbf{M}\cdot \mathbf{Q} \cdot \mathbf{M} \label{MatrixEq} \; ,\ee
which is linear in $\mathbf{Q}$.
For $0<p_z\leq 1$ it has the unique solution 
\be \mathbf{Q} = \frac{p_z +1}{2p_z} \mathbf{P} \,+\, \frac{p_z-1}{2p_z} \mathbf{M}\cdot \mathbf{P} \cdot \mathbf{M} \label{Solution} \; .\ee
Since the matrices $Q$ and $P$ coincide with the states $\sigma^{kk'}_\mathcal{E'}$ and $\sigma^{kk'}_\mathcal{E}$ obtained via the Jamiolkowski isomorphism \cite{Isomorphism}
 (see Sec.\ref{upper1}), the conditions that $\mathbf{Q}$ actually corresponds to a completely positive and trace preserving map $\mathcal{E'}$, are
 (a) $\text{tr}_{k'}\, \mathbf{Q} = \sigma_0^k$ and (b) $\mathbf{Q} > 0$, i.e. that $\mathbf{Q}$ is a density matrix.
It is straightforward to show that $\text{tr}_{k'}\, \mathbf{Q}$ is the maximally mixed state $\sigma_0=\frac{1}{2}\mathbf{1}$ on particle $k$ whenever $\text{tr}_{k'}\, \mathbf{P} = \sigma_0^k$ and hence that for $0<p_z\leq 1$ condition (a) is always fulfilled. On the other hand the positivity (b) imposes further constraints on the allowed parameter range of $p_z$, for which the channel 
$\mathcal{E}$ permits a decomposition of the form (\ref{Decomp}).
For Pauli channels $ \mathcal{E}\rho = \sum_{i=0,1,2,3} \, p_{i} \, \sigma_i \rho \sigma_i $ the solution $\mathbf{Q}$ is again a diagonal matrix:
\bea \mathbf{Q} & = & \frac{1}{2p_z}\text{diag}\, \big (\,(p_z+1)p_0 + (p_z-1)p_3, \cr 
& & (p_z+1)p_1 +  (p_z-1)p_2,\cr 
& &(p_z+1)p_2 + (p_z-1)p_1, \cr 
& & (p_z+1)p_3 + (p_z-1)p_0\,\big ) \; . \label{PauliSol}\eea
Now the positivity (b) yields the four inequalities:
\bea 
\frac{1-p_z}{1+p_z} \, \leq\, \frac{p_0}{p_3} \, \leq\, \frac{1+p_z}{1-p_z} \; , \cr
\frac{1-p_z}{1+p_z} \, \leq\, \frac{p_1}{p_2}\, \leq\, \frac{1+p_z}{1-p_z} \; ,
\eea 
which imply that in particular 
 either both $p_1$ and $p_2$ [$p_0$ and $p_3$] are zero or none of them. 
Therefore in the case of a bitflip channel, as to be expected, no dephasing component can be extracted in the above sense.
Moreover it is straightforward to see that the minimal value for $p_z$, that allows for such an extraction is given by 
\be \frac{1-p_z}{1+p_z} \leq q_{\text{min}} \equiv \min \left\{ \frac{p_0}{p_3}, \frac{p_1}{p_2},\frac{p_3}{p_0},\frac{p_2}{p_1}\right\} \; ,\ee i.e.
  $p_z \geq \frac{1-q_{\text{min}}}{1+q_{\text{min}}} $.
In the trivial case of a dephasing channel itself with parameter $p$, i.e. $p_0=\frac{1+p}{2}$, $p_3=\frac{1-p}{2}$ and $p_1=p_2=0$, 
this clearly gives a minimal value for $p_z=p$. For the depolarizing channel ($p_0=\frac{1+3p}{4}$, $p_1=p_2=p_3=\frac{1-p}{4}$) we obtain the minimal value for
 $p_z=\frac{2p}{1+p}$, and in the case of the quantum optical channel with $\mu=0$ ($p_0 = 1 -(2p+q)$, $p=1=p_2=p\leq\frac{1}{4}$ and $p_3=q\leq\frac{1}{4}$ )
 one arrives at $p_z= 1 - \frac{q}{1-2p}$. 

\section*{Appendix B}
In this appendix we will give the proof to the statements $(i),(ii)$ and $(iii)$ of Sec.\ref{upper3}.

{\it Proof of (i):}
Let $U_1,U_2,U_3$ denote the disjoint subsets on which a $\sigma_x,\sigma_y$ or $\sigma_z$ error occurs. From the stabilizer equations
 $K^G_k |G\rangle=|G\rangle$ it
follows that $\sigma_x^k|G\rangle=\sigma_z^{N_k}|G\rangle$ and similarly $\sigma_y^k|G\rangle=\sigma_z^k\,\sigma_x^a|G\rangle=\sigma_z^{N_k+k}|G\rangle$.
With this one obtains
\begin{eqnarray}  
\rho  & = & \; \sum_{U_1,U_2,U_3\subseteq V \atop U_i\cap U_j = 0} \, p_0^{|V|-|U_1|-|U_2|-|U_3|}\, p_1^{|U_1|}\, p_2^{|U_2|}\,p_3^{|U_3|} \cr
 & & \, \sigma_z^{U} |G \rangle \langle G | \sigma_z^{U}\, ,
\end{eqnarray} 
where $U=\Gamma(U_1+U_2) +U_2 +U_3$.
 With $q_i=\frac{p_i}{p_0}$ the eigenvalues $\lambda_U$ therefore can be written as 
\[
\lambda_U   = p_0^{|V|}\, \sum_{(U_1,U_2,U_3) \in \mathcal{M}(U)} \, q_1^{|U_1|}\, q_2^{|U_2|}\,q_3^{|U_3|}\; ,
\]
where 
\begin{eqnarray}
\mathcal{M}(U) & = &\{\, (U_1,U_2,U_3)\; |\; U_i \subseteq V\,, \, U_i\cap U_j =0 ,\cr 
 & & \; \; U=\Gamma(U_1+U_2) +U_2 +U_3\,\} \; .\nonumber 
\end{eqnarray}
 This set may alternatively be written as 
\begin{eqnarray}  
\mathcal{M}(U) & = & \{\; \left( U'\setminus U'' \,,\, U'\cap U''\, ,\, U''\setminus U' \right) \cr 
& &\;\; |\; U',U''\subseteq V \, ,\, U = \Gamma U' + U''\;\} \cr
 & = & \big\{\; \big(U'\setminus (\Gamma U' +U)\,,\,U'\cap (\Gamma U' +U)\,,\cr
& & \; \; (\Gamma U' +U)\setminus U'\big) \;|\; U'\subseteq V \;\big\} 
\; . \nonumber
\end{eqnarray} 
This is exactly the index set in Eq.~(\ref{PauliLambda}).

{\it Proof of (ii):}
We first note that for partial transposition $^{T_A}$ generally 
\begin{eqnarray}\label{PTrule}
& &\left(C^{A}_1\otimes C^{B}_2\, {\bf D}^{AB} \, C^{A}_3\otimes C^{B}_4 \right)^{T_A} \cr
& = & (C^{T}_3)^A\otimes C^{B}_2\, ({\bf D}^{AB})^{T_A} \, (C^{T}_1)^A \otimes C^{B}_4
\end{eqnarray}
 holds. With this rule we can compute $$ K^G_k \left(|G \rangle\langle G| \right)^{T_A} K^G_k = \left( K^G_k |G \rangle\langle G| 
 K^G_k \right)^{T_A} = \left(|G \rangle\langle G| \right)^{T_A}\; ,$$
  i.e. $\left(|G \rangle\langle G| \right)^{T_A}$ commutes 
 with $ K^G_k $ for all $k \in V$. Therefore $\left(|G \rangle\langle G| \right)^{T_A}$ is again diagonal in the graph state basis
 $\{ |U\rangle_G \, | \, U\subseteq V\} $, i.e.
 $ \left(|G \rangle\langle G| \right)^{T_A} = \sum_{U \subseteq V} \, \lambda'_U\,  |U \rangle_G\langle U|$ .
 In order to determine the spectrum let us decompose $U$ into $(U_A, U_{A^c})$ according to the partitioning and use the Schmidt decomposition \cite{He03} 
 \[ |G \rangle = \frac{1}{2^{\frac{|A|}{2}}} \, \sum_{A'\subseteq A} \, (-1)^{f_A(A')} \, |A'\rangle^A |\Gamma' A'\rangle^{A^c}\]
 with $f_A(A') = \langle A', \Gamma_A A' \rangle$, the standard $z$-basis $|C\rangle^A = \sigma_x^C |0\rangle^{\otimes A}$ on partition $A$ and the graph basis $|C\rangle^{A^c} = |C\rangle_{G\setminus A}^{\otimes A^c}$ on partition $A^c$ with respect to the pure graph state corresponding to the graph
 $G\setminus A$, which is obtained from $G$ by removing all vertices in $A$. Now we can compute
 \begin{eqnarray}
 \lambda'_U & := & _G\hspace{-0.05cm}\langle U |  \left(|G \rangle\langle G| \right)^{T_A}  | U \rangle_G \nonumber \\
 &  = & \frac{1}{4^{|A|}} \sum_{A_i \subseteq A \atop i=1,2,3,4} (-1)^{\sum_{i} f_A(A_i) + \langle U_A, A_1 +A_2 \rangle} \nonumber \\
 & & \langle A_1 | A_3 \rangle \langle A_2 | A_4 \rangle \langle \Gamma' A_1 | \Gamma'A_2 + U_{A^c} \rangle \times \nonumber \\
& & \hspace{1cm} \times  \langle \Gamma' A_3 + U_{A^c} | \Gamma'A_4  \rangle \nonumber \\
\vspace{0.2cm}
 & = & \frac{1}{4^{|A|}} \sum_{A_1,A_2 \subseteq A : \atop \Gamma' (A_1 + A_2) = U_{A^c}} (-1)^{\langle U_A, A_1 + A_2\rangle} \nonumber \\
 & = & \frac{1}{2^{|A|}} \sum_{A' \subseteq A : \atop \Gamma' A' = U_{A^c}} (-1)^{\langle U_A, A'\rangle} 
 \; .  \label{eqend}
\end{eqnarray}
 Whereas $\lambda_U$ vanishes for $U\subseteq V$ with $U_{A^c} \notin \text{Im}\, \Gamma' $, in the opposite case $U_{A^c} \in \text{Im}\, \Gamma' $
 we can choose an arbitrary $A_0$ with $\Gamma' A_0 = U_{A^c}$ to simplify (\ref{eqend}):  
\begin{eqnarray}
 \lambda'_U  & = & \frac{1}{2^{|A|}}  (-1)^{\langle U_A, A_0 \rangle}  \sum_{A' \in \text{ker}\, \Gamma'} (-1)^{\langle U_A, A'\rangle} \cr 
 & = & \frac{1}{8^{|A|}}  (-1)^{\langle U_A, A_0 \rangle}   \left \{
\begin{array}{ccc}
|\text{ker}\, \Gamma'| & \text{if}&  U_A \in  (\text{ker}\, \Gamma')^{\bot},\\
0 & \text{if} &  U_A \notin  (\text{ker}\, \Gamma')^{\bot}\\
\end{array} 
\right .  \nonumber
\end{eqnarray}
Note that this is independent of the choice $A_0$, since a different choice $\tilde{A_0}$ will differ from $A_0$ only by an element $A' \in \text{ker}\, \Gamma'$,
for which $\langle U_A, A' \rangle = 0$ if $U_A \in  (\text{ker}\, \Gamma')^{\bot}$.
Therefore one obtains the partial transpose for the pure graph state
\begin{eqnarray}
 \label{PTofG}
(|G \rangle\langle G|)^{T_A}  = \frac{|\text{ker}\,\Gamma'|}{2^{|A|}} \hspace{1cm} \times \hspace{2cm} & & \cr
\sum_{(X,Y) \in \atop (\text{ker}\,\Gamma')^{\bot} \times (\text{Im}\,\Gamma')}  
(-1)^{\langle X, A_Y \rangle}   | X + Y \rangle_G\langle X + Y|    & & 
\end{eqnarray}
The corresponding formula $(\ref{PTofRhoG})$ for a general graph diagonal state can be deduced from Eq.~(\ref{PTofG}) by again using $(\ref{PTrule})$.\\

{\it Proof of (iii):}
For $\lambda_{U+k}$ the estimation can be derived from Eq. $(\ref{PauliLambda})$:
By adding [deleting] an element $k$ to [from] the set $\Gamma U' + U$ the corresponding
sizes of the sets $U_1= U'\setminus (\Gamma U' +U)$, $U_2 = U'\,\cap\, (\Gamma U' +U)$ and $U_3=(\Gamma U' +U)\setminus U'$ in the exponents of $(\ref{PauliLambda})$
can at most increase or decrease by one. Moreover, since $U_1, U_2$ and $U_3$ are disjoint, the operation $U \mapsto U+k$ of adding or deleting $k$
cannot simultaneously increase or decrease any two sets $U_i$. For example, if $k \in U_1$ then 
\begin{eqnarray}
| U'\setminus(\Gamma U' + U + k)| & = & |U_1| - 1 \; ,\cr 
|U'\cap (\Gamma U' + U + k)| &  = & |U_2| + 1 \; , \cr  
|(\Gamma U' +U)\setminus U'| & = &|U_3| \; . \nonumber
\end{eqnarray} 
In any case every addend 
\[ q_1^{|U'-(\Gamma U' + U +k)|}\, q_2^{|U'\cap (\Gamma U' + U + k)|} \, q_3^{| (\Gamma U' +U +k)-U'|} \]
  in $(\ref{PauliLambda})$ can be bounded from below by $q \times  \, q_1^{|U_1|}\, q_2^{|U_2|}\,q_3^{|U_3|}$ and from above by
  $\frac{1}{q} \times \, q_1^{|U_1|}\, q_2^{|U_2|}\,q_3^{|U_3|}$ since $q\leq 1$. This gives $q \lambda_U\leq \lambda_{U+k}\leq \frac{1}{q}\lambda_U $.
For $\lambda_{U+N_k}$ and $\lambda_{U+N_k+k}$  a similar argument holds, if one rewrites 
\begin{eqnarray}
\lambda_{U+N_k} & = &  p_0^{|V|} \,  \sum_{U'\subseteq V} \, q_1^{|(U'+k)\setminus (\Gamma U' +U)|} \, \times \cr
& &  \times \, q_2^{|(U'+k)\cap (\Gamma U' +U)|}  \, q_3^{| (\Gamma U' +U)\setminus (U'+k)|}\nonumber
\end{eqnarray}
and
\begin{eqnarray} 
\lambda_{U+N_k+k} & = &  p_0^{|V|} \,  \sum_{U'\subseteq V} \, q_1^{|(U'+k)\setminus (\Gamma U' + U +k)|}\, \times \cr 
& &  \times \, q_2^{|(U'+k)\cap (\Gamma U' + U +k)|}  \, q_3^{| (\Gamma U' + U +k)\setminus (U'+k)|} \; . \nonumber
\end{eqnarray}
In this representation we 'absorbed' $N_k$ into the summation index $U'$ in both cases using $\Gamma U' + U + N_k =\Gamma'(U'+k) + U$. This concludes the proof of (iii).





\begin{thebibliography}{99}

\bibitem{Fo00}
S. Braunstein and Hoi-Kwong Lo, Fortschritte der Physik, Vol. {\bf 48}, Number 9-11 (2000).

\bibitem{Sr35} E. Schr\"{o}dinger, Die Naturwissenschaften {\bf 23}, 807 (1935).

\bibitem{Zu03}
For a recent review see e.g. W. Zurek, quant-ph/0306072.

\bibitem{Du04b}
W. D\"ur and H.-J. Briegel, Phys. Rev. Lett. {\bf 92}, 180403 (2004) 

\bibitem{Rau01}
H.J. Briegel and R. Raussendorf, Phys. Rev. Lett. {\bf 86}, 910 (2001).

\bibitem{He03}
M. Hein, J. Eisert and H.-J. Briegel, Phys. Rev A {\bf 69}, 062311 (2004).

\bibitem{Shor} P.W.Shor, J. Math. Phys. Vol. {\bf 43}, 4334-4340 (2002).

\bibitem{Br93} H.J. Briegel and B.G. Englert, Phys. Rev. A {\bf 47}, 3311 (1993);
H.J. Briegel, Ph.D. thesis, LMU M\"unchen, Germany (1993).

\bibitem{multiparty} W. D\"ur and J. I. Cirac, Journal of Physics A: Mathematical and General, Vol. {\bf 34}, No. 35, 6837-6850 (2001).

\bibitem{Th02}
A. V. Thapliyal and J. A. Smolin, quant-ph/0212098.

\bibitem{Du00}
W. D\"ur and J. I. Cirac, Phys. Rev. A {\bf 61}, 042314 (2000); 
W. D\"ur and J. I. Cirac, Phys. Rev. A {\bf 62}, 022302 (2000). 

\bibitem{Ho97}
M. Horodecki, P. Horodecki and R. Horodecki, 
Phys. Rev. Lett. {\bf 78}, 574 (1997).

\bibitem{Pe96}
A. Peres, Phys. Rev. Lett. {\bf 77}, 1413 (1996).

\bibitem{Ho96} M. Horodecki, P. Horodecki and
R. Horodecki, Phys. Lett. A{\bf 223}, 8 (1996).

\bibitem{BI}  N. Gisin and H. Bechmann-Pasquinucci, Phys. Lett. A {\bf 246}, 1-6 (1998).

\bibitem{Secure}
M. Hillery, V. Buzek and A. Berthiaume, Phys. Rev. A {\bf 59}, 1829 (1999);
D. Gottesmann, Phys. Rev. A {\bf 61}, 042311 (2000);
C.Crepeau, D. Gottesmann and A.Smith, Proc.34th ACM STOC, 643-652, New York (2002).

\bibitem{Frequency}
S.F. Huelga, C. Macchiavello, T. Pellizzari, A.K. Ekert, M.B. Plenio and J.I.Cirac,
Phys. Rev. Lett. {\bf 79}, 3865 (1997).

\bibitem{Si02}
C. Simon and J. Kempe, Phys. Rev. A {\bf 65}, 052327 (2002).

\bibitem{graphCodes} 
D. Schlingemann and R.F. Werner, Phys. Rev. A {\bf 65}, 012308 (2001);
D. Schlingemann, Quant. Inf. Comp. 2, No {\bf 4}, 307-323, (2002).

\bibitem{Lo04} 
K. Chen and H. Lo, quant-ph/0404133; E. Rains, private communications.

\bibitem{oneWay}
H.-J. Briegel and R. Raussendorf, Phys. Rev. Lett. {\bf 86}, 910 (2001);
R. Raussendorf, D. Browne and H.J. Briegel, Phys. Rev. A {\bf 68},022312 (2003).
 

\bibitem{Realisation}
D. Jaksch, H.J. Briegel, J.I. Cirac, C.W. Gardiner and P.Zoller, Phys. Rev. Lett. {\bf 82}, 1975 (1999);
M. Greiner, O. Mandel, T.W. H\"ansch, and I. Bloch, Nature {\bf 419}, 51-54 (2002).

\bibitem{Ma03}
M. Van den Nest, J. Dehaene, and B. De Moor, Phys. Rev. A {\bf 69}, 022316 (2004),
quant-ph/0405023 and quant-ph/0404106.

\bibitem{Du04}
W. D\"ur, H. Aschauer, H.-J. Briegel, Phys. Rev. Lett. {\bf 91}, 107903 (2003);
H. Aschauer, W. D\"ur and H.-J. Briegel, quant-ph/0405045. 

\bibitem{notedegree}
This condition is in fact too stringent, as full distillability of the $N$--particle state $\rho$ already follows from distillability of all such pairs of particles $(k,l)$ which form a fully connected graph $G$. 

\bibitem{Isomorphism}
A. Jamiolkowski, Rep. Mod. Phys. {\bf 3}, 275-278 (1972).

\bibitem{Ho98}
M. Horodecki, P. Horodecki and R. Horodecki, 
Phys. Rev. Lett. {\bf 80}, 5239 (1998).

\bibitem{Ci00}
J. I. Cirac, W. D\"ur, B. Kraus and M. Lewenstein, 
Phys. Rev. Lett. {\bf 86}, 544 (2001);
W. D\"ur and J. I. Cirac, Phys. Rev. A {\bf 64}, 012317 (2001).

\bibitem{St98}
A. Steane, `Quantum Error Correction' in {\em Introduction to Quantum Computation and Information}, 
edited by H-K. Lo, T. Spiller and S. Popescu, World Scientific 1998. 





\end{thebibliography}
\end{document}